\documentclass{aa}
\usepackage[T1]{fontenc}
\usepackage[varg]{txfonts}
\usepackage{multirow}
\usepackage{color}
\usepackage{times}
\usepackage{amssymb}
\usepackage{amsmath}
\usepackage{booktabs}
\usepackage{upgreek}
\usepackage{natbib}
\usepackage{array}
 \newcolumntype{C}[1]{>{\centering\let\newline\\\arraybackslash\hspace{0pt}}m{#1}}
 \newcolumntype{R}[1]{>{\raggedleft\let\newline\\\arraybackslash\hspace{0pt}}m{#1}}

 \newcommand{\OIII}{\ion{O}{iii}}

 \newcommand{\NII}{\ion{N}{ii}}
 \newcommand{\SII}{\ion{S}{ii}}

 \newcommand{\Hb}{\mbox{H$\beta$}}
 \newcommand{\Ha}{\mbox{H$\alpha$}}

\bibpunct{(}{)}{;}{a}{}{,}

\begin{document}

\title{Revisiting dual AGN candidates with spatially resolved LBT spectroscopy\thanks{The LBT is an international collaboration among institutions in the United States, Italy and Germany. LBT Corporation partners are: The University of Arizona on behalf of the Arizona Board of Regents; Istituto Nazionale di Astrofisica, Italy; LBT Beteiligungsgesellschaft, Germany, representing the Max-Planck Society, The Leibniz Institute for Astrophysics Potsdam, and Heidelberg University; The Ohio State University, and The Research Corporation, on behalf of The University of Notre Dame, University of Minnesota and University of Virginia.}}
\subtitle{The impact of spillover light contamination}

\author{B.~Husemann\inst{1}
\and
J.~Heidt\inst{2}
\and
A.~De Rosa\inst{3}
\and
C.~Vignali\inst{4,5}
\and
S.~Bianchi\inst{6}
\and
T.~Bogdanovi\'c\inst{7}
\and
S.~Komossa\inst{8}
\and
Z.~Paragi\inst{9}
}
\institute{
Max-Planck-Institut f\"ur Astronomie, K\"onigstuhl 17, D-69117 Heidelberg, Germany\\
\email{husemann@mpia.de}
\and
Landessternwarte, Zentrum f\"ur Astronomie der Universit\"at Heidelberg, K\"onigstuhl 12, 69117 Heidelberg, Germany
\and
INAF – Istituto di Astrofisica e Planetologie Spaziali, Via Fosso del Cavaliere, 00133 Rome, Italy
\and
Dipartimento di Fisica e Astronomia, Alma Mater Studiorum, Università degli Studi di Bologna, Via Gobetti 93/2,40129 Bologna, Italy
\and
INAF - Osservatorio di Astrofisica e Scienza dello Spazio di Bologna, Via Gobetti 93/3, I-40129 Bologna, Italy
\and
Dipartimento di Matematica e Fisica, Universit\`a degli Studi Roma Tre, via della Vasca Navale 84, I-00146 Roma, Italy
\and 
Center for Relativistic Astrophysics, School of Physics, Georgia Institute of Technology, 837 State Street, Atlanta, GA 30332-0430
\and
Max-Planck-Institut f{\"u}r Radioastronomie, Auf dem H{\"u}gel 69, 53121 Bonn, Germany
\and Joint Institute for VLBI ERIC (JIVE),
Oude Hoogeveensedijk 4, 7991PD Dwingeloo, Netherlands
}

\authorrunning{Husemann et al.}
\titlerunning{Revisiting dual AGN candidates with spatially resolved LBT spectroscopy}
\abstract{The merging of supermassive black holes (SMBHs) is a direct consequence of our hierarchical picture of galaxy evolution. It is difficult to track the merging process of SMBHs during mergers of galaxies as SMBHs are naturally difficult to observe.}
{We want to characterise and confirm the presence of two independent active galactic nuclei (AGN) separated by a few kiloparsec in seven strongly interacting galaxies previously selected from the Sloan Digital Sky Survey (SDSS) as Seyfert-Seyfert pairs based on emission-line ratio diagnostics.}
{Optical slit spectra taken with MODS at the Large Binocular Telescope (LBT) are presented to infer the detailed spatial distribution of optical emission lines, and their line ratios and AGN signatures with respect to the host galaxies, thereby  quantifying the impact of beam smearing and large fibre apertures on the spectra captured by the SDSS.}
{We find that at most two of the seven targets actually retain a Seyfert-Seyfert dual AGN, whereas the others may be more likely powered by post-AGB stars in retired galaxies or through shocks in the ISM based on spatially resolved optical line diagnostics. The major cause of this discrepancy is a bias caused by the spillover of flux from the primary source in the secondary SDSS fibre which can be more than an order of magnitude at $<$3\arcsec\ separations. Previously reported extremely low X-ray--to--[\ion{O}{iii}] luminosity ratios may be explained by this misclassification, as can heavily obscured AGN for the primaries. We also find that the nuclei with younger stellar ages host the primary AGN.}
{Studies of close dual AGN selected solely from fibre-based spectroscopy can create severe biases in the sample selection and interpretation of the results. Spatially resolved spectroscopy should ideally be used in the future to characterise such compact systems together with multi-wavelength follow-up observations.}
\date{}{}
\maketitle

\section{Introduction}
If galaxies merge hierarchically and most galactic bulges contain supermassive black holes (SMBHs), the formation of dual or multiple SMBHs is a natural phenomenon \citep{Begelman:1980}. Multiple mergers offer a potential physical mechanism linking star formation on a galaxy-wide scale with the feeding and evolution of active galactic nuclei (AGN) \citep[e.g.][]{Hopkins:2008a}. The evolution of merging SMBH systems from several kiloparsec to smaller separations is determined by gravitational interactions of the SMBHs with their environment \citep{Mayer:2013}, mainly dynamical friction \citep[e.g.][]{Escala:2004}, and the scattering of the SMBHs by massive gas clouds and spiral arms produced during the merger phase \citep{Fiacconi:2013}. 

Active galactic nuclei trace the active easily observable phase of SMBHs, and are therefore the ideal objects where  SMBHs can be discovered. Since dual SMBHs (i.e. with a separation of 100\,pc up to 10\,kpc) are a natural consequence of galaxy mergers, their search received great attention \citep[see][and references therein]{Bogdanovic:2009,deRosa:2019}. Evidence of such systems has proven elusive until now. In particular, only a small number of dual AGN have been successfully identified at the centres of single-host galaxies so far. Some of the clearest examples are  NGC~6240 \citep{Komossa:2003},
0402+379 \citep{Rodriguez:2006}, SDSS~J1536$+$0441 \citep{Bondi:2010}, LBQS~0103$-$2753 \citep{Shields:2012}, Mrk~463 \citep{Bianchi:2008}, and  SDSS~J1323$-$0159 \citep{Woo:2014}. However, the number of confirmed dual or multiple AGN remains small, and advances in this field have been meager so far.

Large fibre-based optical surveys such as the Sloan Digital Sky Survey \citep[SDSS,][]{York:2000} and the Large Sky Area Multi-Object Fibre Spectroscopic Telescope \citep[LAMOST,][]{Cui:2012} have enabled large systematic searches for close dual AGN candidates selected as double-peaked emitters in the [\ion{O}{iii}] $\lambda\lambda$4960,5007 emission \citep{Wang:2009,Liu:2010a, Smith:2010,Ge:2012,Barrows:2013,Shi:2014,Wang:2019}. The assumption is that double-peaked [\ion{O}{iii}] emitters are potentially caused by the presence of two independent AGN narrow-line regions (NLRs) around distinct AGN because of the large velocity offset required   to produce two line peaks (several 100\,km/s). A few hundred double-peaked [\ion{O}{iii}] emitters have been discovered this way, which represent roughly 1\%\ of the screened parent AGN population. However, dual AGN are not the only explanation for the double-peaked [\ion{O}{iii}] emission. Bi-polar AGN-driven outflows \citep{Rosario:2010b,Shen:2011b,Fu:2012,MuellerSanchez:2015,McGurk:2015,Nevin:2016}, rotating gas discs \citep{Smith:2012}, or a single AGN photo-ionising the interstellar media of {both} galaxies in a  merger \citep{Xu:2009} can also explain the line shapes, which emphasises the need for spatially resolved spectroscopic follow-up of each candidate to understand the exact nature of the kinematics. 

A much more robust detection of dual AGN candidates is expected if the nuclei exhibit AGN signatures from distinct spectra rather than double-peaked emission in a single spectrum. Such a catalogue  of projected AGN pairs in the range from 5 to 100\,kpc was established by \citet{Liu:2011} from the SDSS data release 7 \citep{Abazajian:2009}. They reported that 3.6\% of AGN are projected and that 30\% of the pairs show morphological disturbances. However, \citet{Hou:2019} was able to confirm two of the five targets from this catalogue as dual AGN based on X-ray follow-up observation with \textit{Chandra}, while the non-detection would imply extreme X-ray--to--[\ion{O}{iii}] flux ratios if they are still genuine dual AGN. 

In this paper we present spatially resolved optical long-slit spectroscopy with the Large Binocular Telescope (LBT) of seven dual AGN with angular separations of $<$9\arcsec\ from the \citet{Liu:2011} catalogue. We specifically explore here how light spillover between distinct SDSS fibres  can arise at close separations due to the relatively large fibre diameters (3\arcsec), poor seeing, and positional uncertainties, which could artificially boost the observed emission line fluxes of a secondary nucleus. These observational issues could naturally lead to the reported extreme X-ray--to--[\ion{O}{iii}] flux ratios as the [\ion{O}{iii}] flux of the putative secondary nuclei could be significantly overestimated for such misclassified dual AGN at small separations.

Throughout the paper we assume a concordance cosmological model with $H_0=70\,\mathrm{km}\,\mathrm{s}^{-1}\,\mathrm{Mpc}^{-1}$, $\Omega_{\mathrm{m}}=0.3$, and $\Omega_\Lambda=0.7$.

\begin{figure*}
\centering
 \includegraphics[width=0.95\textwidth]{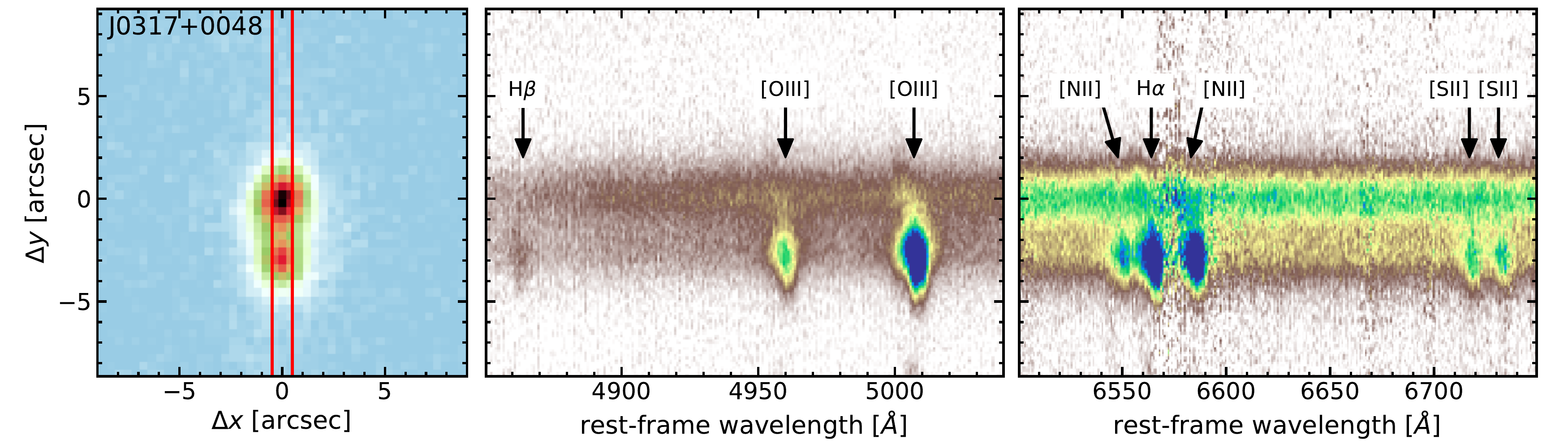}\\
 \includegraphics[width=0.95\textwidth]{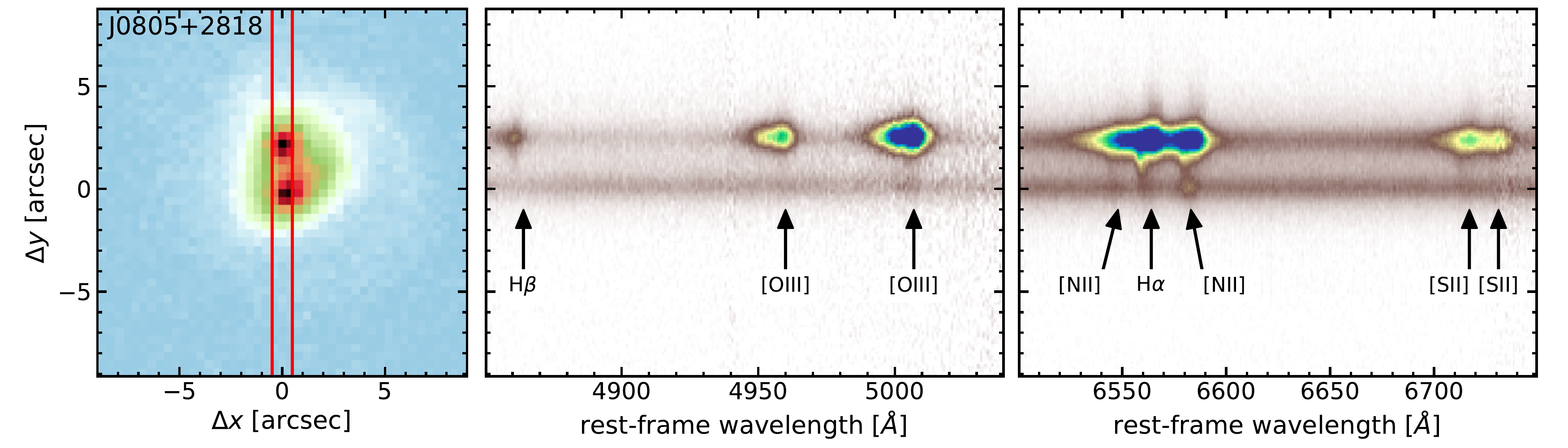}\\
 \includegraphics[width=0.95\textwidth]{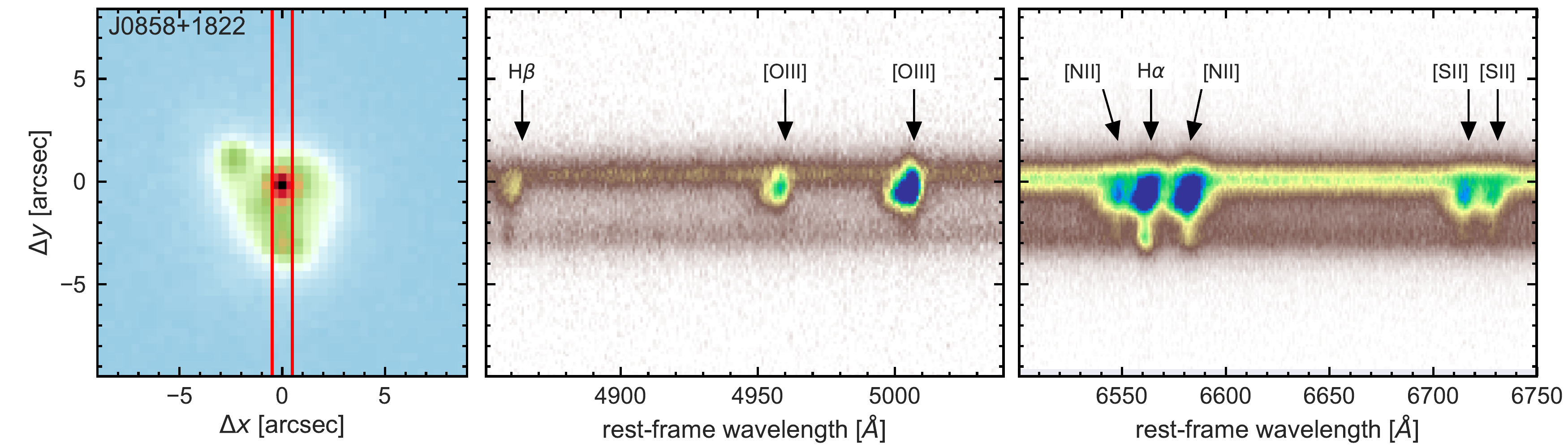}\\
 \includegraphics[width=0.95\textwidth]{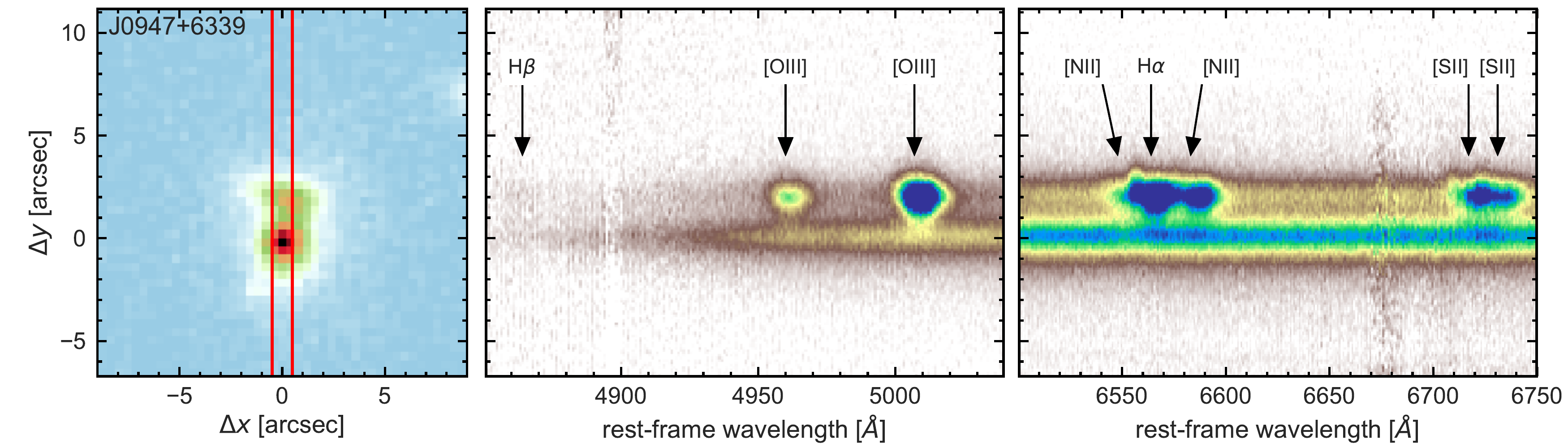}\\
 \caption{Overview of LBT slit observations for each target. \textit{Left panel:} SDSS $r$-band image of the source rotated such that the slit is orientated in the vertical direction, as indicated by the two red lines. \textit{Right panels:} Cutout of the LBT MODS 2D spectra focused on the wavelength region covering the \Hb/[\OIII] and the \Ha/[\NII]/[\SII] region. The emission lines are labelled. }\label{fig:2Dspec}
 \end{figure*}
 \setcounter{figure}{\value{figure}-1}
 \begin{figure*}
 \includegraphics[width=0.95\textwidth]{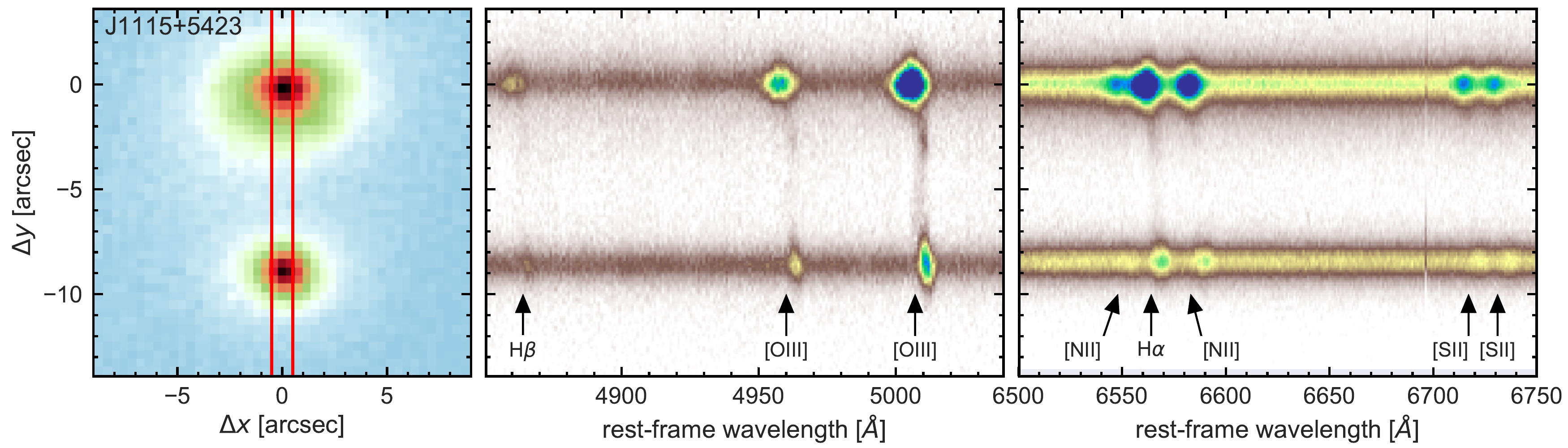}\\
 \includegraphics[width=0.95\textwidth]{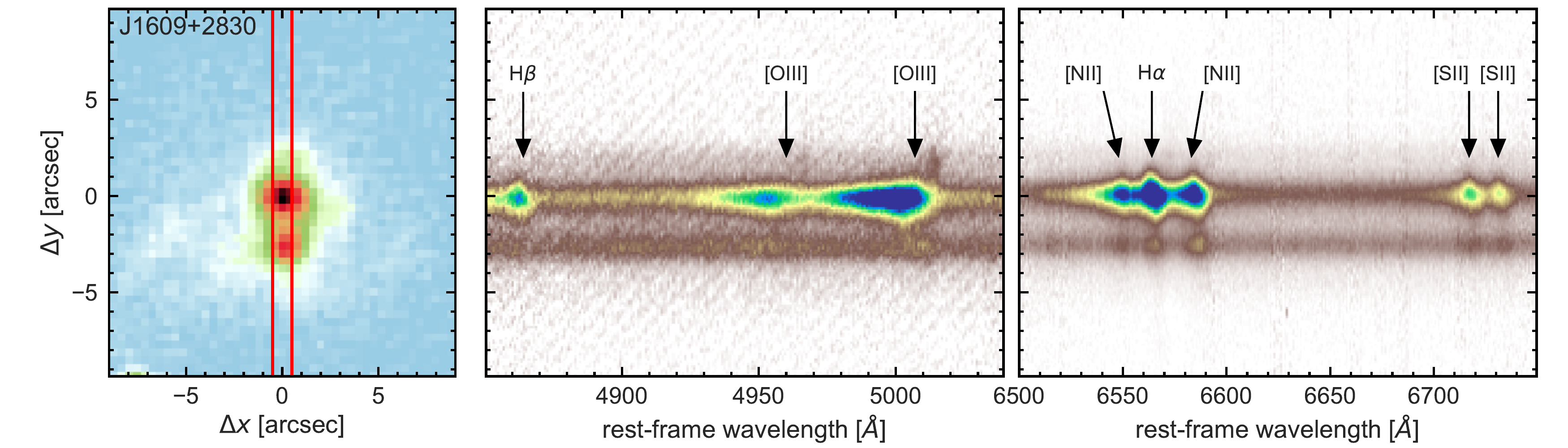}\\
 \includegraphics[width=0.95\textwidth]{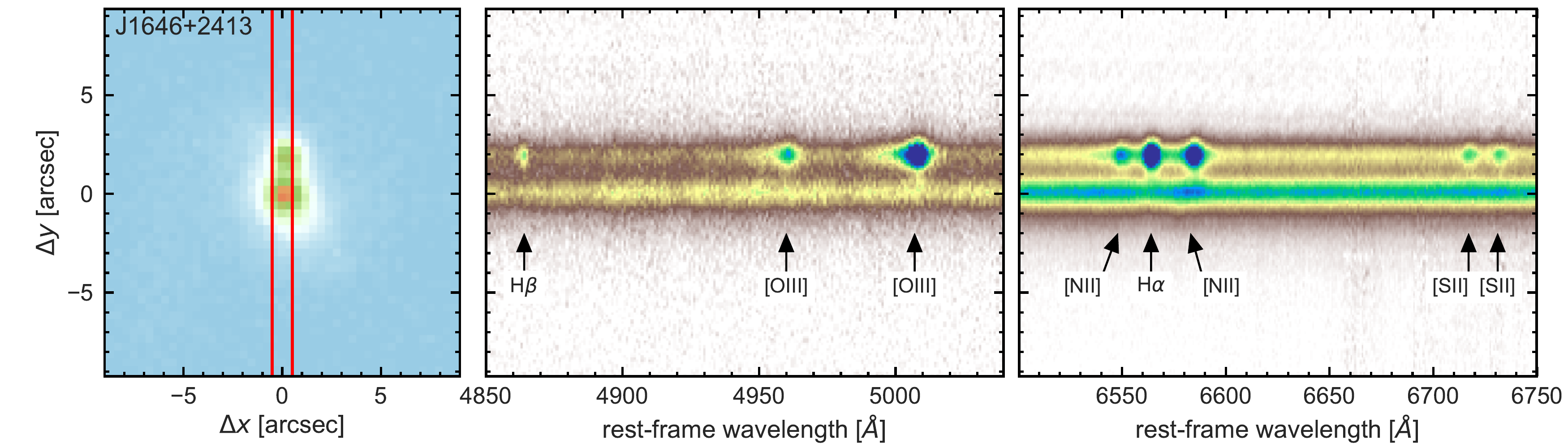}
 \caption{continued}
\end{figure*}

\begin{table*}
\caption{Sample and observations}\label{tab:observations}
\centering
\begin{tabular}{lccccccc}\hline\hline
  Name\tablefootmark{a}                            &  $z$      & $\Delta d$\tablefootmark{b} & instrument & date       &  $t_\mathrm{exp}$  & airmass & seeing\tablefootmark{c}\\\hline
  SDSS J031722.06$+$004801.8      &  0.15722  & 2\farcs8   & MODS1      & 2016-02-05 &  $2\times750$\,sec &   1.5   & 2\farcs0\\ 
  SDSS J080523.29$+$281815.8      &  0.12844  & 2\farcs3   & MODS1      & 2016-02-04 &  $2\times450$\,sec &   1.1   & 1\farcs3\\ 
  SDSS J085837.68$+$182223.4      &  0.05894  & 2\farcs6   & MODS1      & 2016-02-04 &  $2\times300$\,sec &   1.1   & 1\farcs1\\ 
  SDSS J094741.58$+$633939.2      &  0.13973  & 2\farcs2   & MODS1      & 2016-02-05 &  $2\times300$\,sec &   1.2   & 1\farcs2\\  
  SDSS J111519.98$+$542316.7      &  0.07043  & 8\farcs3   & MODS1      & 2016-02-04 &  $2\times300$\,sec &   1.1   & 1\farcs1\\ 
  SDSS J160933.41$+$283058.4      &  0.16960  & 2\farcs7   & MODS2      & 2016-05-26 &  $3\times900$\,sec &   1.2   & 1\farcs2\\
  SDSS J164658.48$+$241134.1      &  0.08728  & 2\farcs1   & MODS2      & 2016-05-26 &  $2\times750$\,sec &   1.2   & 1\farcs1\\\hline 
 \end{tabular}
 \tablefoot{
 \tablefoottext{a}{Identifier from the Sloan Digital Sky Survey of one of the nuclei.}
 \tablefoottext{b}{Angular distance between the two nuclei as inferred from the optical continuum light.}
 \tablefoottext{c}{Inferred FWHM of the seeing during the observations at 6000\,\AA.}
 }
\end{table*}

\section{Small-separation dual AGN candidates}\label{sect:sample}
\subsection{Sample selection}
The Sloan Digital Sky Survey \citep[SDSS,][]{York:2000} provides a large collection of galaxy spectra from which close pairs of AGN can be drawn. However, the tiling of the plates requires a minimum distance of 55\arcsec\ between the fibres \citep{Blanton:2003b}. Hence, spectroscopy of two targets with smaller separations can only be obtained if neighbouring fibre-plate fields are overlapping on the sky or if repeated observations of the same field are taken with different fibre positions for small-separation targets. Mining  SDSS data release 7 \citep[DR7,][]{Abazajian:2009}, \citet{Liu:2011} collected a sample of 1286 candidate multiple AGN systems with physical separations $<$100\,kpc from SDSS-DR7 in the redshift range $0.02<z<0.33$. They selected obscured  (narrow-line) AGN following an emission-line diagnostics classification  \citep{Kauffmann:2003} based on MPA-JHU value-added emission-line catalogues \citep{Brinchmann:2004,Tremonti:2004}. They further complemented the sample with narrow-line quasars from \citet{Reyes:2008} and unobscured (broad-line) AGN from \citet{Hao:2005}, and \citet{Schneider:2010} within the same redshift range as the narrow-line AGN.

Here, we focus on narrow-line AGN pairs where emission-line diagnostic clearly identified Seyfert-like ionisation, and we exclude pairs with LINER-like emission following the demarcation line of \citet{Stasinska:2008}. Since LINERs can also be powered by other mechanisms than AGN photo-ionisation, such as  shocks \citep[e.g.][]{Heckman:1980} or post-AGB stars \citep[e.g.][]{Singh:2013}, we want to avoid this additional confusion in the ionisation mechanisms and select a clean obscured AGN pair sample based on the recorded fibre spectra. Excluding as well  the very wide pairs with  $>$60\,kpc separation, which are not necessarily bound systems, leads to a  subsample of only 17 potential Seyfert-Seyfert pairs.  We selected 6  of the 17 pairs with separations smaller than $3\arcsec$ corresponding to $<$10\,kpc separation at the corresponding redshifts. All these dual AGN candidates are clearly associated with interacting systems that display at least two independent nuclei and tidal features (see Fig.~\ref{fig:2Dspec}). In addition, we selected one target at a slightly larger separation of $\sim$8\arcsec\ in a pair of galaxies as a control galaxy for our study. The selected sample and the separations of the apparent nuclei are listed in Table~\ref{tab:observations}.

\begin{figure}
 \resizebox{\hsize}{!}{\includegraphics{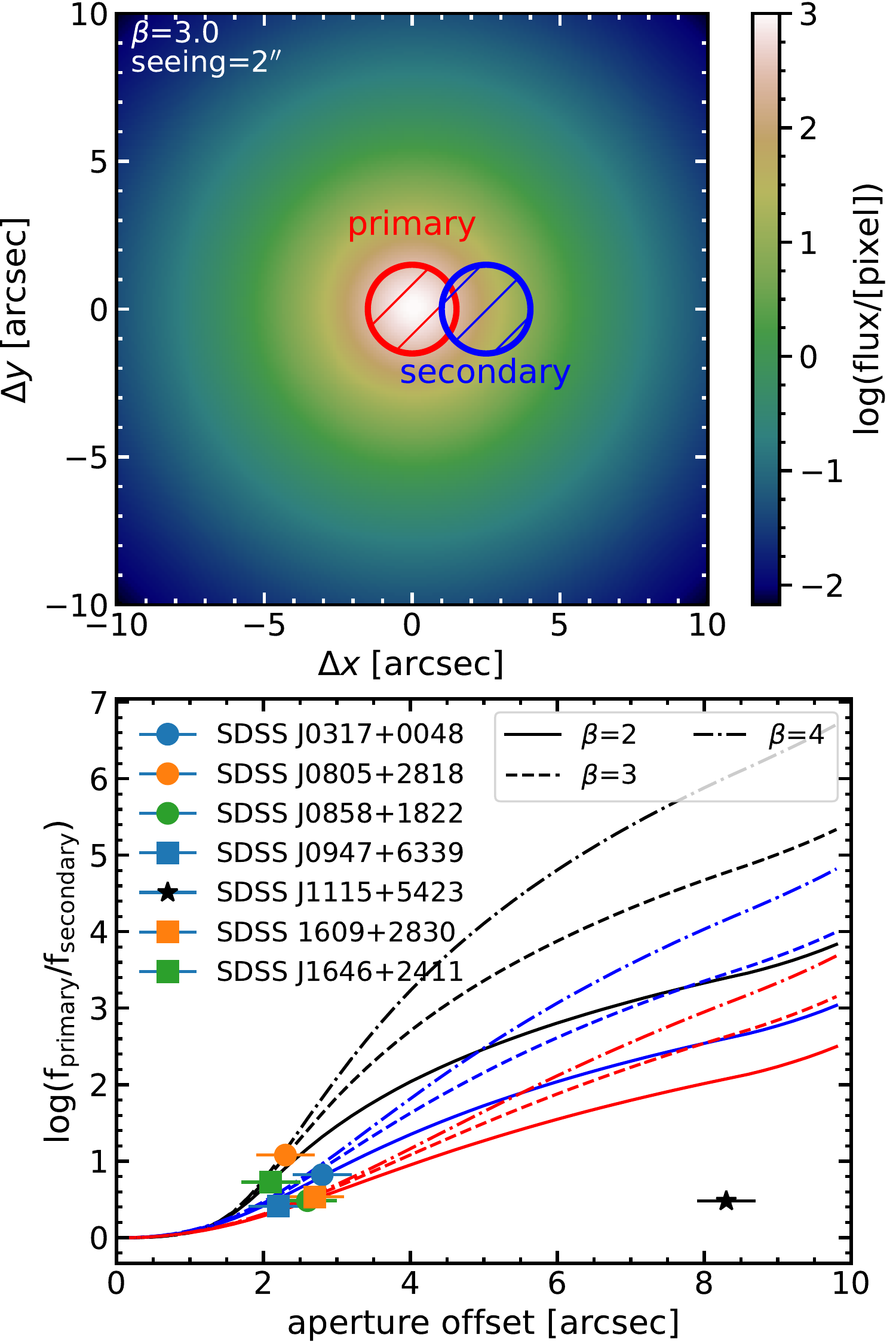}}
 \caption{Theoretical expectations for flux ratio between primary and offset secondary SDSS fibre in the case of a point source. \textit{Top panel:} Logarithmic light distribution of a point source assuming a circular 2D Moffat function with $\beta=3$ and seeing of 2\arcsec. The circles indicate the two independent SDSS fibres with a diameter of 3\arcsec\ where the primary is centred on the target and the secondary is offset by 2\farcs5. \textit{Bottom panel:} Predictions for the flux ratios of the primary and secondary fibre for a point source as a function of secondary fibre distance to the point source. The curves are computed for three different seeing values of 1\arcsec\ (black curves), 2\arcsec\ (blue curves), and 3\arcsec\ (red curves) with three different $beta$ values each as described in the legend. The measured [\ion{O}{iii}] $\lambda5007$ flux ratios for the six targets are shown as symbols;    the error bar indicates the error on the combined fibre position uncertainty of 0\farcs4 for the two fibres.} 
 \label{fig:PSFsimulation}
\end{figure}

\subsection{Predictions of light spillover and line flux biases}
Although a dual AGN is a compelling hypothesis for these sources, two additional possibilities are usually overlooked: (1) only one AGN is ionising gas out to the kpc-scale distances \citep[e.g.][]{Husemann:2014} seen in the secondary fibre, or (2) the large SDSS fibres can be significantly contaminated by the flux from the primary source due to the extended wings of the point-spread function (PSF). In both cases the secondary fibre may contain emission-line ratios consistent with AGN ionisation, but significantly lower in absolute flux than the primary spectrum. In particular the last point is important for very close dual AGN candidates when the separation of the putative nuclei is getting close to the fibre diameters and seeing of the spectroscopic SDSS observations. It is important to note that SDSS spectroscopic observations are often taken during relatively poor seeing $>$1\farcs7 \citep[e.g.][]{Abazajian:2003} as the good seeing conditions were preferentially used for the imaging. 

The relatively poor seeing of the SDSS observations can cause a significant flux spillover of a point source into an offset fibre. The quantitative strength of this spillover can in principle be accurately calculated if the shape of the PSF and the light distribution of the source is accurately known. Unfortunately, this is usually not the case for fibre-based spectroscopic observations, but rough predictions of the effect can be obtained by making reasonable assumptions. Here, we assume a simple circular 2D Moffat function for the PSF
\begin{equation}
 I(r) = I_0\left[1+\left(\frac{r}{\alpha}\right)^2\right]^{-\beta}, \label{eq:moffat}
\end{equation}
where $I_0$ is the peak intensity, $r$ is the radial distance, $\beta$ controls the slope of the wings, and $\alpha$ is a scale factor for the width that scales as $\mathrm{FWHM} = 2\alpha\sqrt{2^{1/\beta}-1}$. The light distribution for $\beta=3$ and a width of $2\arcsec$ (FWHM) is shown in Fig.~\ref{fig:PSFsimulation} (top panel) where the area covered by a central fibre and an offset fibre with a displacement of 2\farcs5 are highlighted for comparison. The corresponding flux ratio $f_\mathrm{primary}/f_\mathrm{secondary}$ as a function of offset distance of the secondary fibre, $\beta$ parameter, and seeing is shown in the bottom panel of Fig.~\ref{fig:PSFsimulation}. 

These predictions are compared against the measured flux ratios of the bright [\OIII]\,$\lambda5007$ line as obtained from the SDSS spectra for our dual AGN candidate sample (Fig.~\ref{fig:PSFsimulation}, bottom panel). Nearly all the objects could in principle be explained entirely by this spillover effect depending on the exact shape of the PSF and fibre position uncertainties during the observations. Only the dual AGN with a separation of more than 5\arcsec\ should not be significantly affected by spillover under realistic conditions. For the others a severe complication with respect to our simple assumption is that these spectra are obtained on different nights and therefore the PSF shape will not necessarily be the same and there is a known uncertainty in the exact position of each fibre of a few 0\farcs1. Hence, it is indispensable to obtain proper spatially resolved spectroscopy of these dual AGN candidates to confirm their original classifications based on the SDSS fibre spectra.

\section{Observations and data reduction}
We obtained spectroscopic observation of the seven candidate dual AGN with the Multi-Object Double Spectrograph \citep[MODS,][]{Pogge:2010} mounted to the Large Binocular Telescope (LBT) at Mount Graham. The data were taken in February and May 2016 using MODS1 and MODS2 in multi-slit mode. Custom masks with 20\arcsec$\times$1\arcsec\ slits were designed to simultaneously observe the primary dual AGN target, some reference stars for simultaneous PSF estimation and sky background together with some filler targets. Observations were split into two or three exposures per mask with a total integration time ranging between 600\,s and 2700\,s. All observations were performed with 1\arcsec\ wide slits in the dual beam G400L/G670L grating mode covering the full optical wavelength range from 3200\AA\ to 10000\AA. The spectral resolution is $R\sim1850$ for the blue channel and $R\sim2300$ for the red channel. Details of the observations are given in Table~\ref{tab:observations}.

We performed the primary detector calibrations such as bias subtraction and flat-fielding with the publicly available python package \textsc{modsCCDRed}\footnote{Available at \url{http://www.astronomy.ohio-state.edu/MODS/Software/modsCCDRed/}}. Afterwards we masked and cleaned cosmic ray hits on individual detector frames with \textsc{PyCosmic} \citep{Husemann:2012a} and processed the science and calibration data with custom-made python scripts. The tracing of the slits along the wavelength axis was achieved through an edge detection in the continuum lamp flat observations, if available, or alternatively using sky background in the science data itself. A wavelength solution was established by tracing various lines of arc lamps along the slit after slit extraction. However, the arc lamps were taken through the wrong slit mask for the target field of SDSS~J0947$+$6333. We traced the wavelength solution through sky lines for the red channel instead, but simply lacked enough bright sky line to derive the solution for the blue channel. Hence, the blue channel data could not be properly reduced for SDSS~J0947$+$6333.

The science spectra were then rectified in spatial and spectral dimension. The background was subtracted by fitting a first-order polynomial in cross-dispersion direction at each wavelength after masking out the target signal. The same processing was also applied to a standard star observations obtained through a fixed long slit with the same slit width to measure the spectrophotometric sensitivity for both spectrograph arms. Finally, we roughly corrected the science spectra for telluric absorption by measuring the strength of the absorption features in the star spectra obtained in the same mask as our target galaxies. 

In Fig.~\ref{fig:2Dspec} we provide an overview of the observation for each dual AGN candidate system. The slits are always oriented such that the two putative AGN within the galaxies are covered simultaneously. The cutouts of the 2D spectra focused on most prominent emission lines in the \Hb/[\OIII] and the \Ha/[\NII]/[\SII] region already reveal very different strength of ionised emission at the two nuclei. Surprisingly, the target SDSS J0858$+$1822 already sticks out from the sample since the light distribution of [\OIII] does not have its peak   at one of the optical continuum peaks of the galaxy.

\begin{figure*}
 \includegraphics[width=0.47\textwidth]{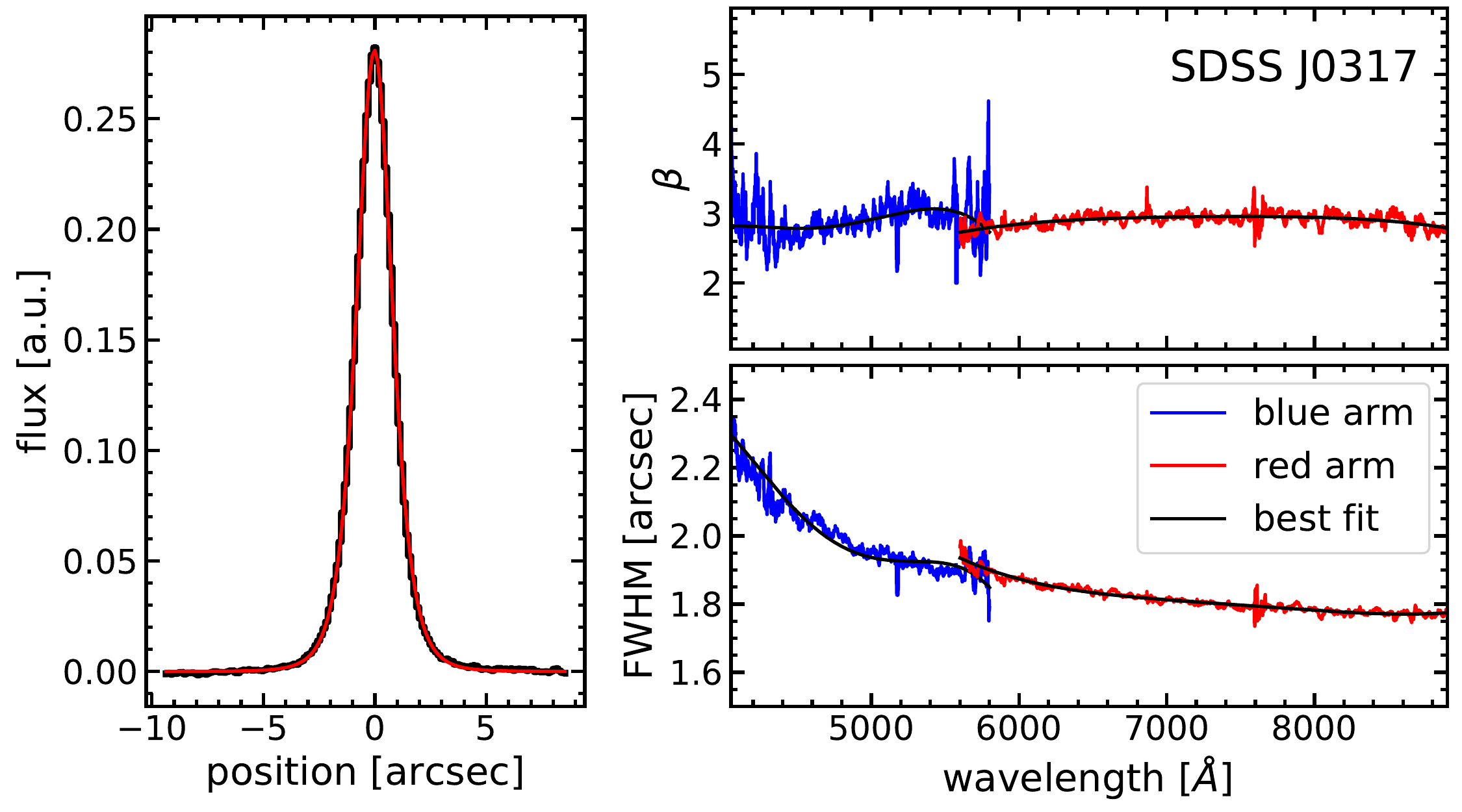}\hfill\includegraphics[width=0.47\textwidth]{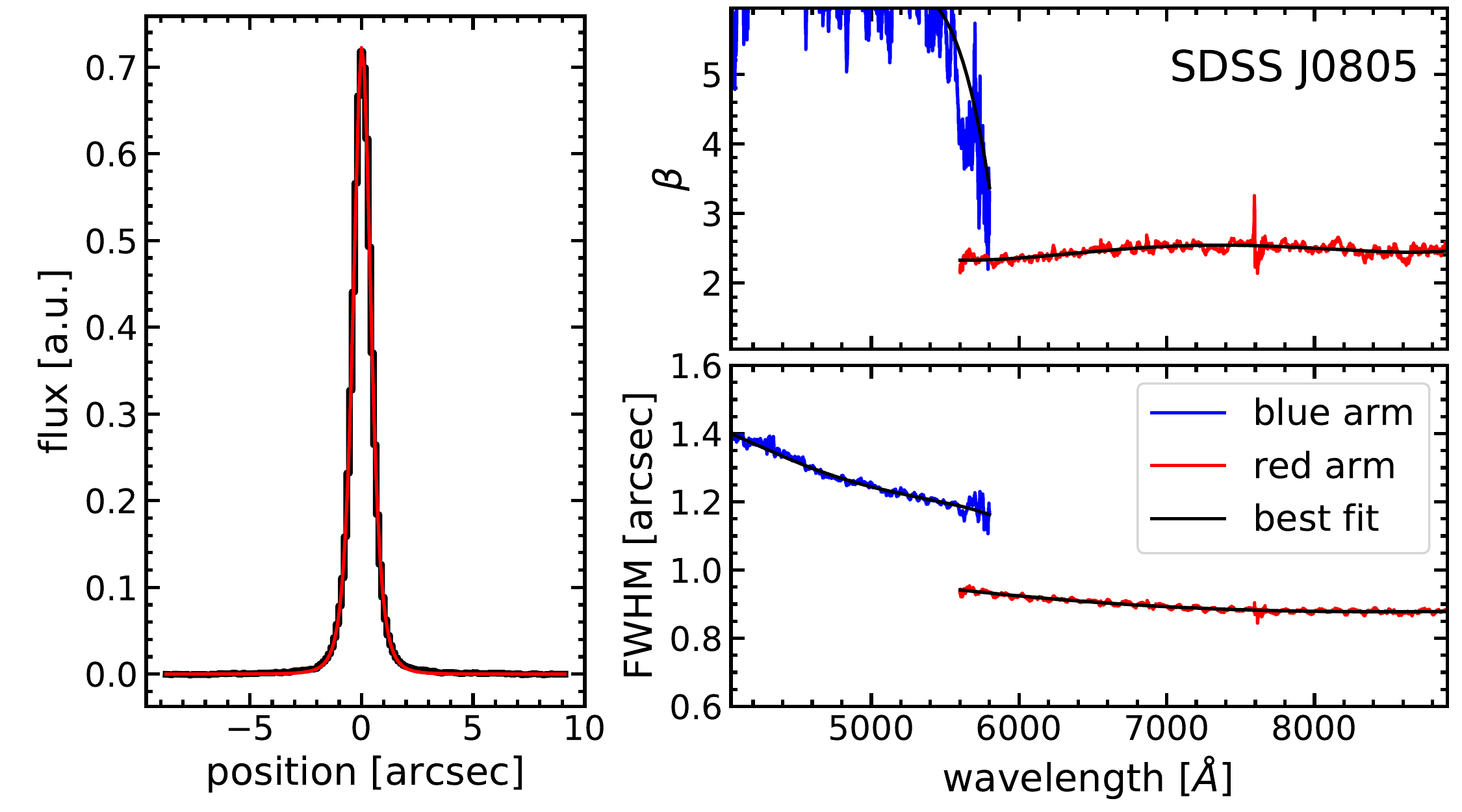}\\
 \includegraphics[width=0.47\textwidth]{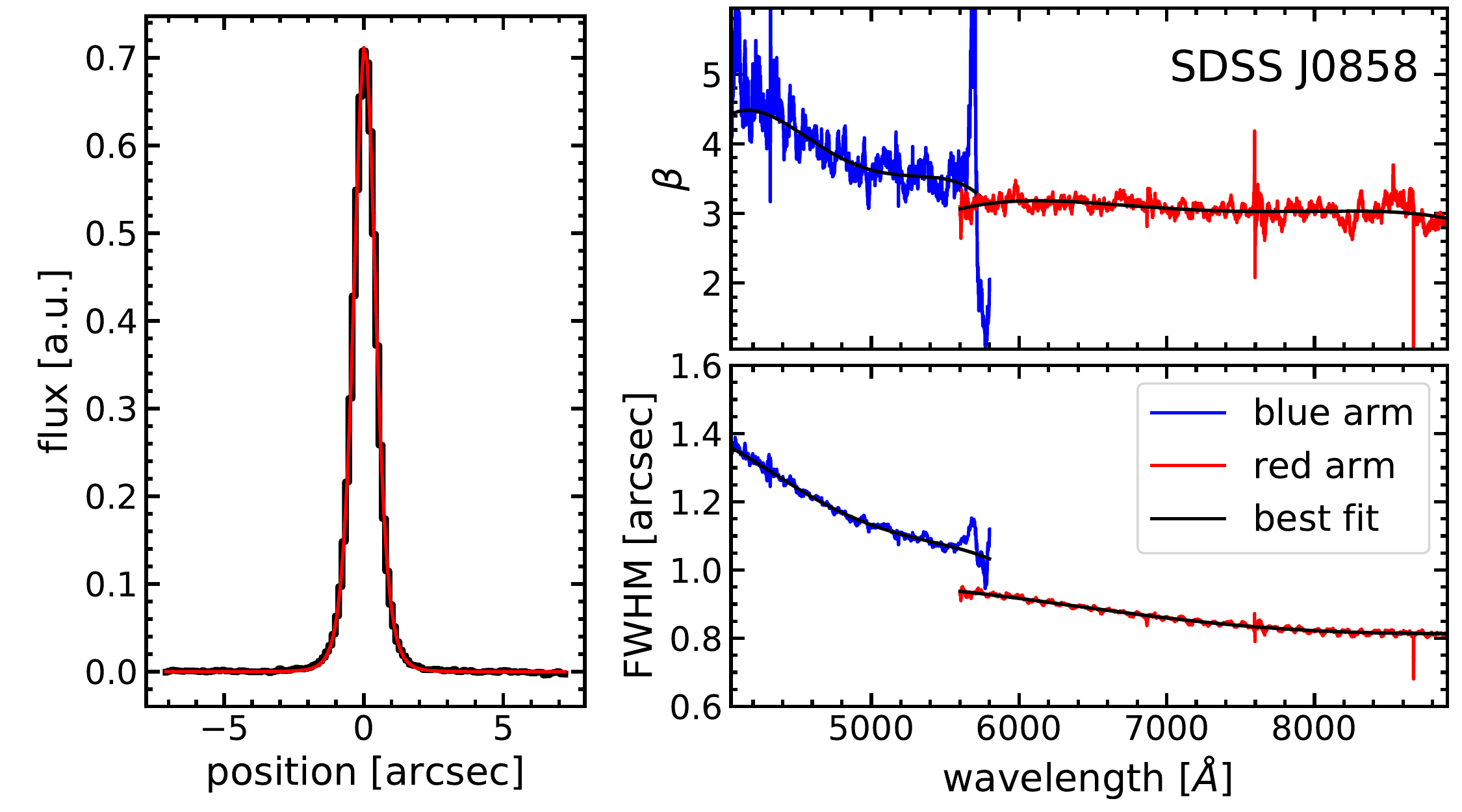}\hfill\includegraphics[width=0.47\textwidth]{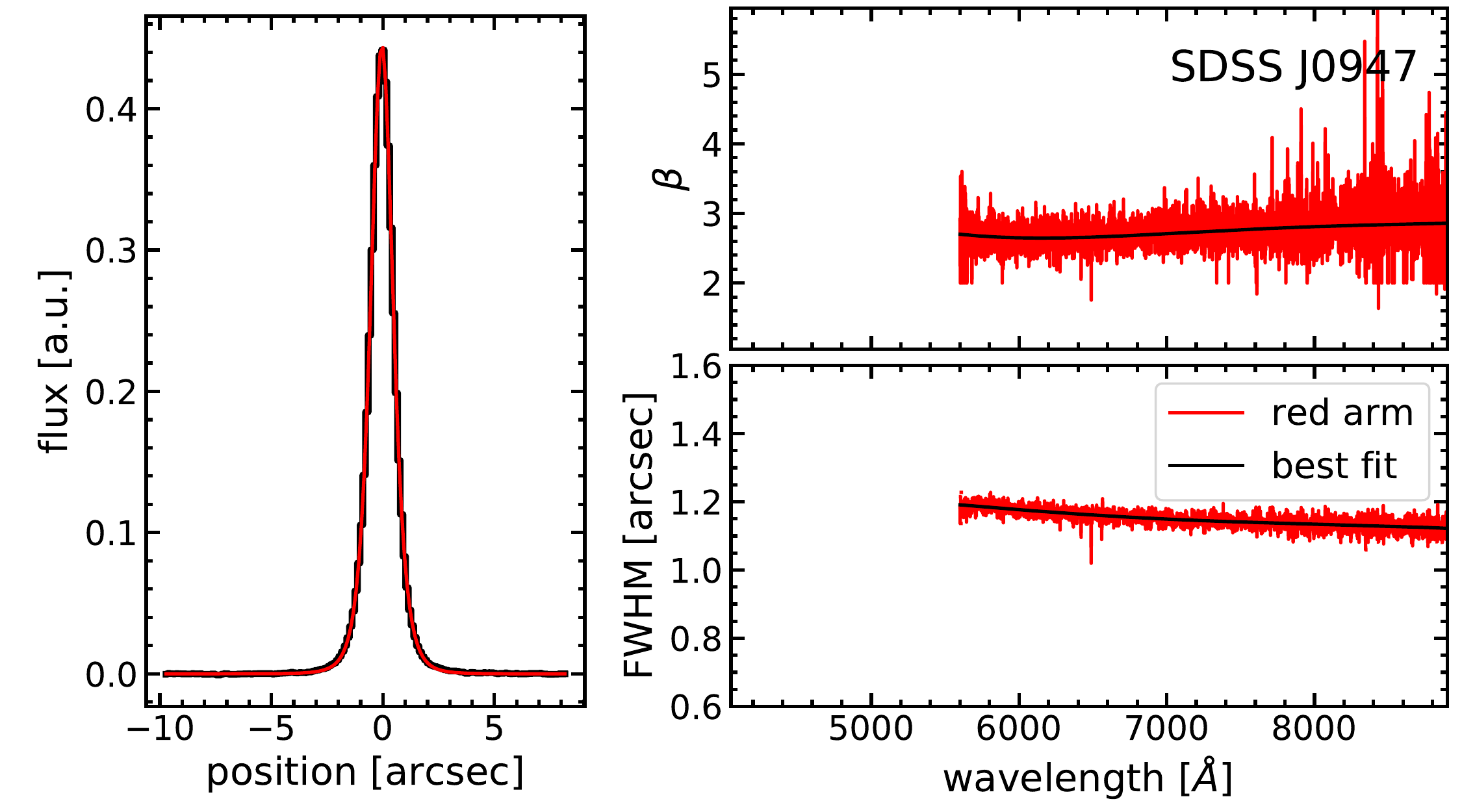}\\
 \includegraphics[width=0.47\textwidth]{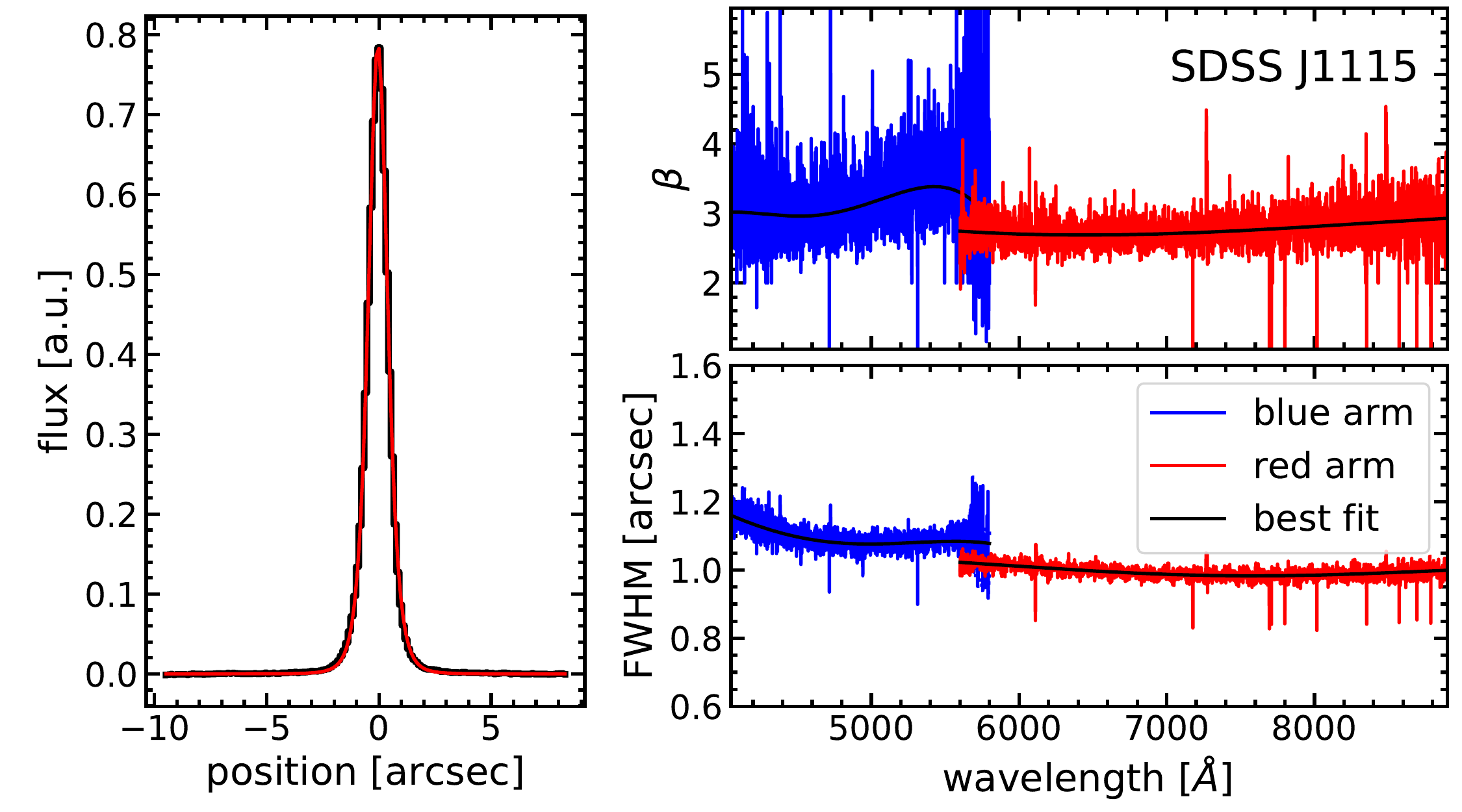}\hfill\includegraphics[width=0.47\textwidth]{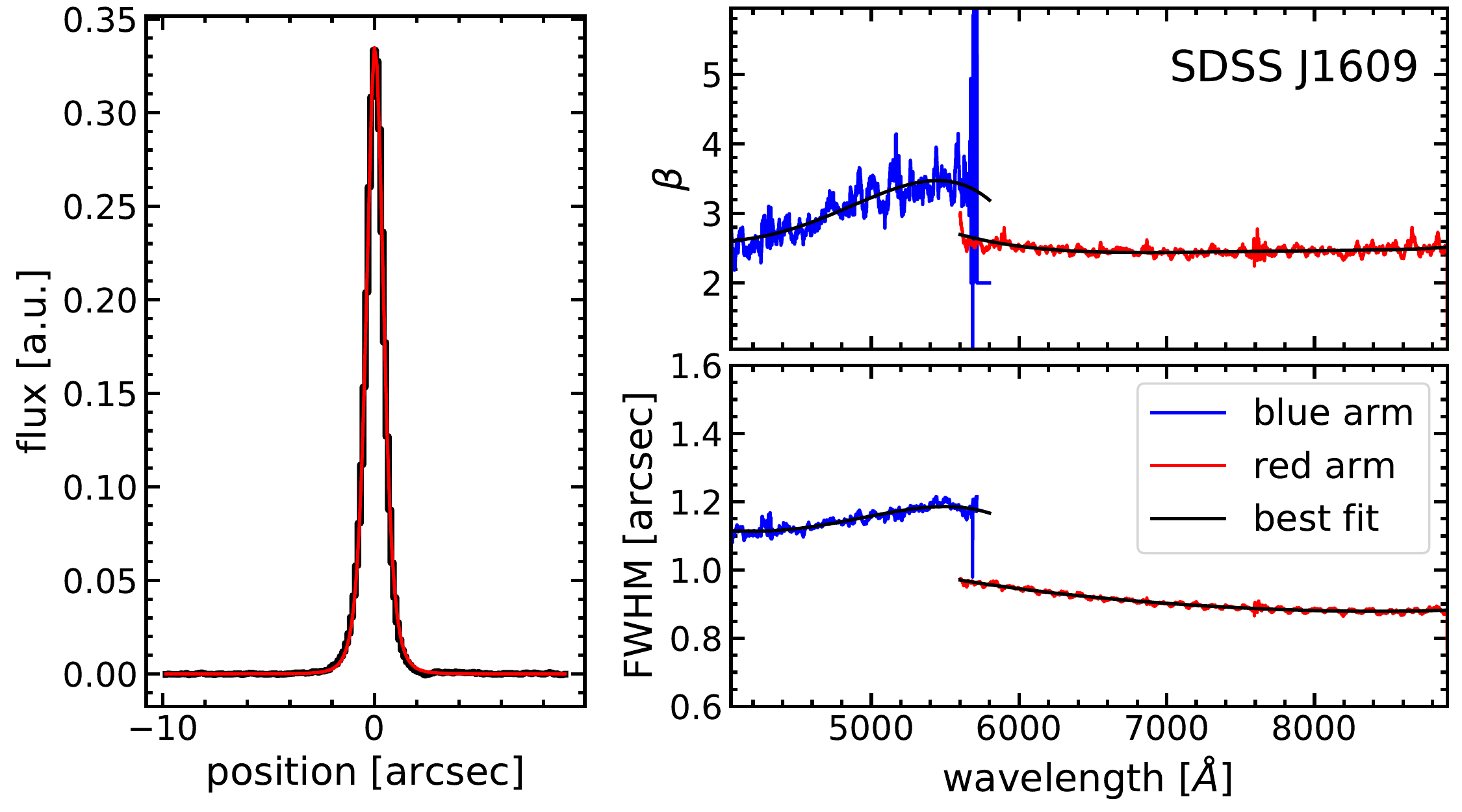}\\
 \centering\includegraphics[width=0.47\textwidth]{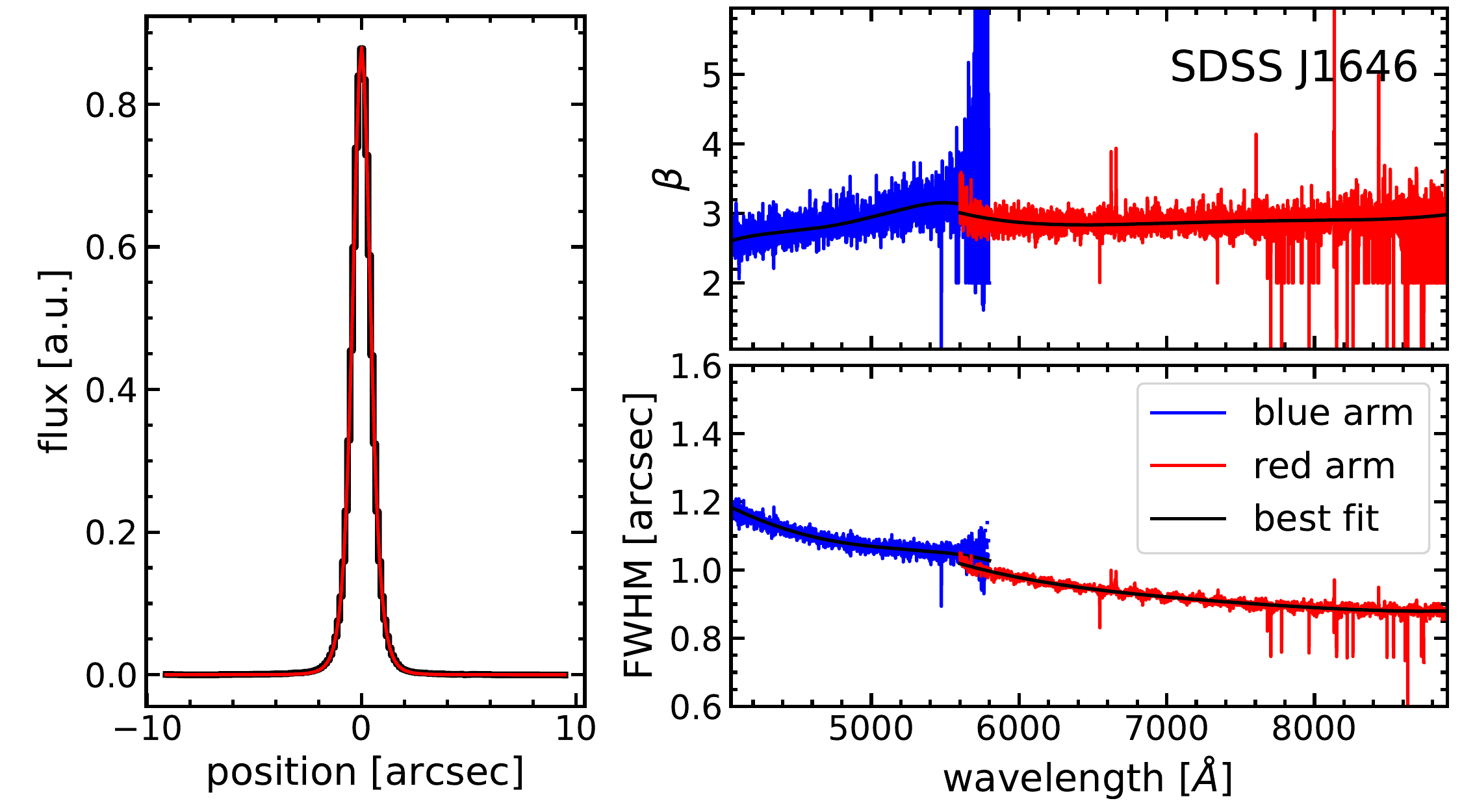}
 \caption{Analysis of the PSF from the star observations in each of the target fields. The cross-dispersion light profile at 6000\AA\ is shown for each star with its best-fit 1D Moffat profile overplotted as the red line in the left panels. The measured $\beta$ and FWHM Moffat shape parameters as a function of wavelength for the red and blue spectrograph arm are shown in the right panels.  A fourth-order polynomial is fitted independently to both channels to smooth the wavelength dependence of the PSF parameters. The obvious breaks in the parameters between the two channels are likely caused by slightly different spectrograph focusing for the different observing nights and instruments.}\label{fig:PSFcharacter}
\end{figure*}

\begin{figure*}
\centering
 \includegraphics[width=0.9\textwidth]{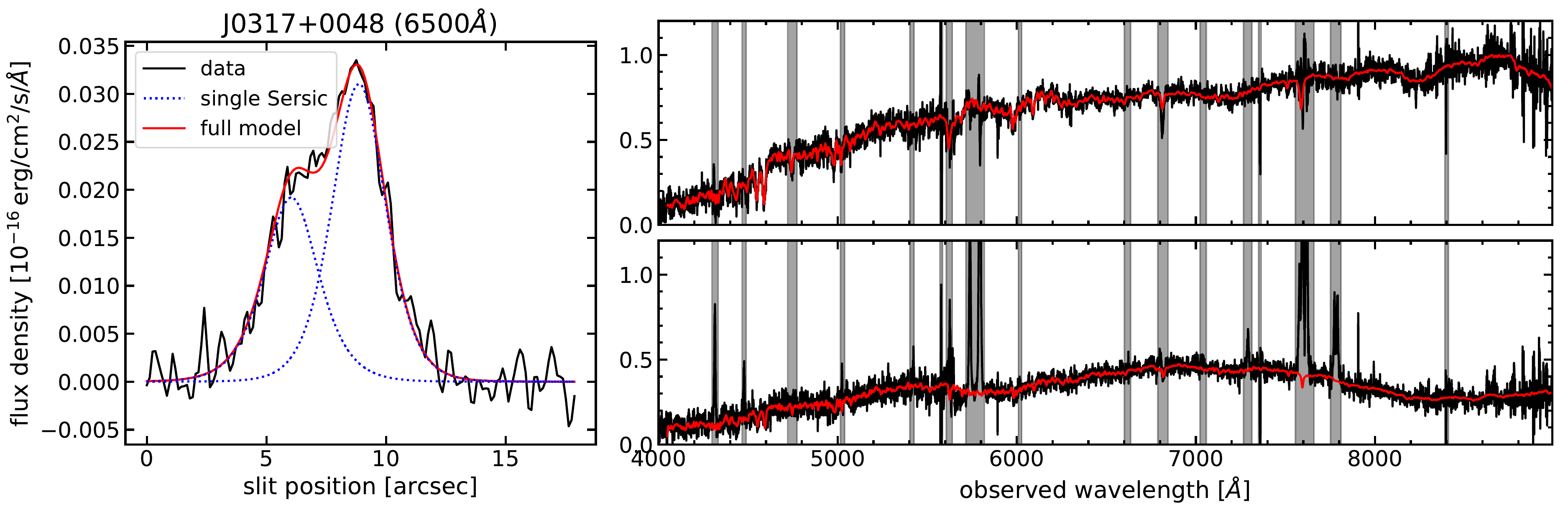}\\
 \includegraphics[width=0.9\textwidth]{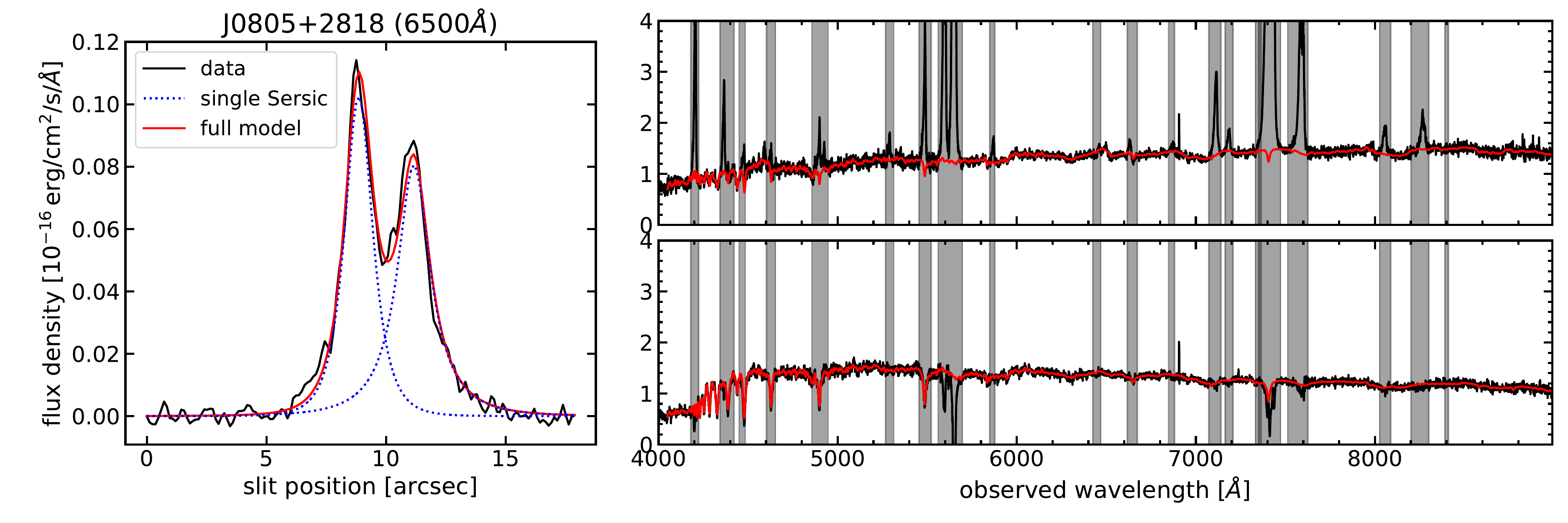}\\
 \includegraphics[width=0.9\textwidth]{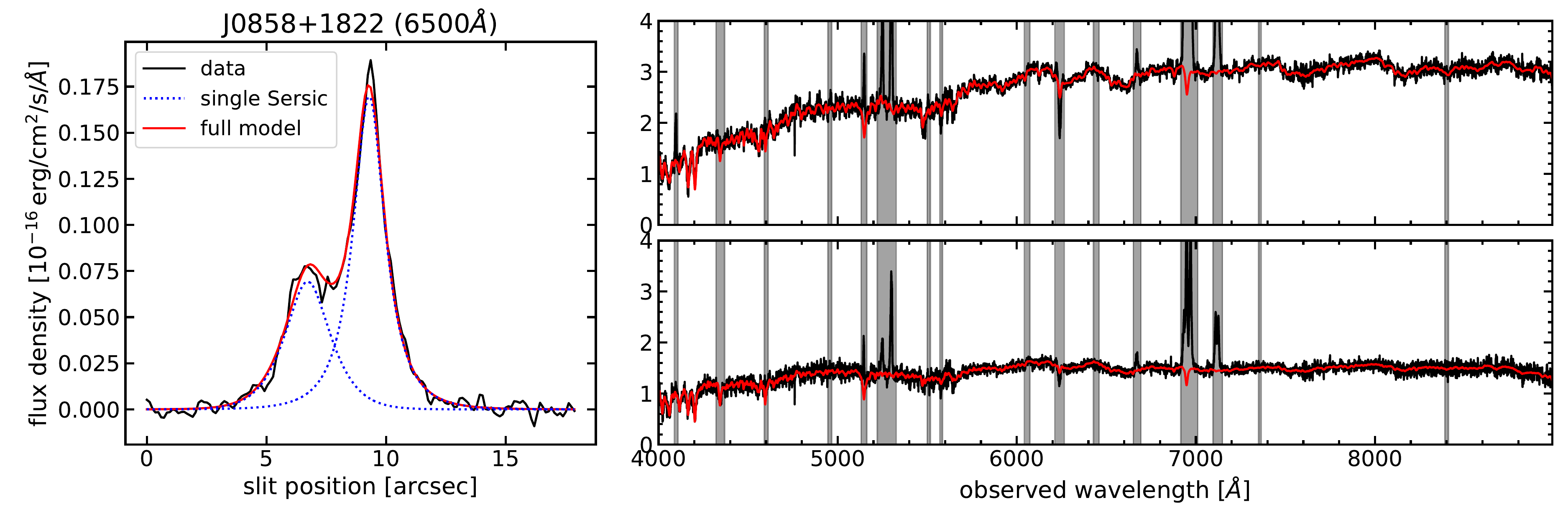}\\
 \includegraphics[width=0.9\textwidth]{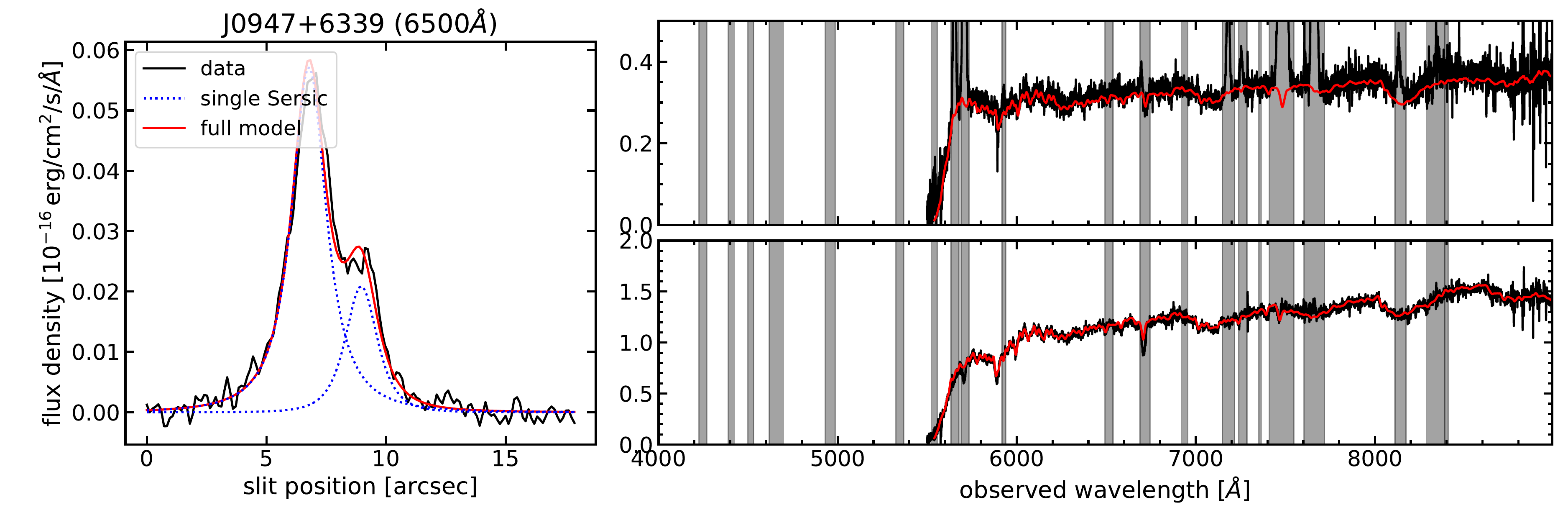}\\
 \caption{Overview of the continuum modelling. \textit{Left panel:} Cross-dispersion profile along the slit at 6500\AA. Our best-fit two-component Sersic model convolved with the PSF is shown for comparison. \textit{Right panel:} Optimally extracted spectra of the two components (black line) assuming that the Sersic parameters only smoothly vary with wavelength. The red line represent the best-fit continuum model based on the INDO-US spectral library fitted with PyParadise. The grey shaded areas indicate the regions masked during the continuum fit.}\label{fig:1Dspec}
\end{figure*}

\setcounter{figure}{\value{figure}-1}
\begin{figure*}
\centering
 \includegraphics[width=0.9\textwidth]{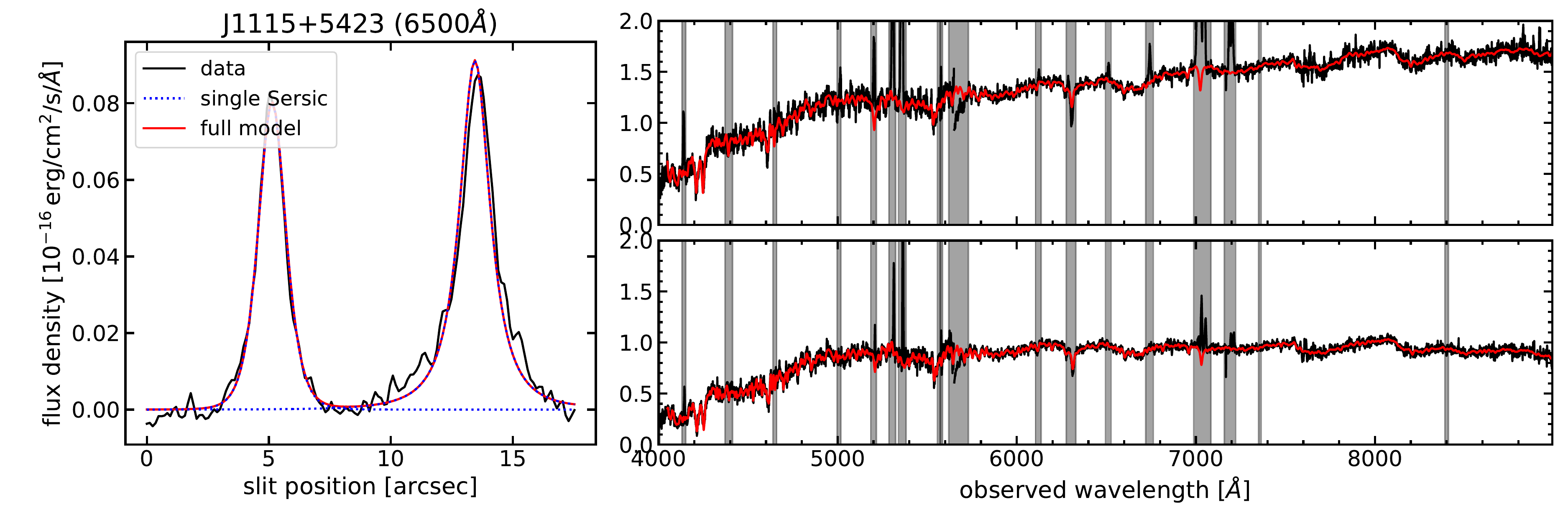}\\
 \includegraphics[width=0.9\textwidth]{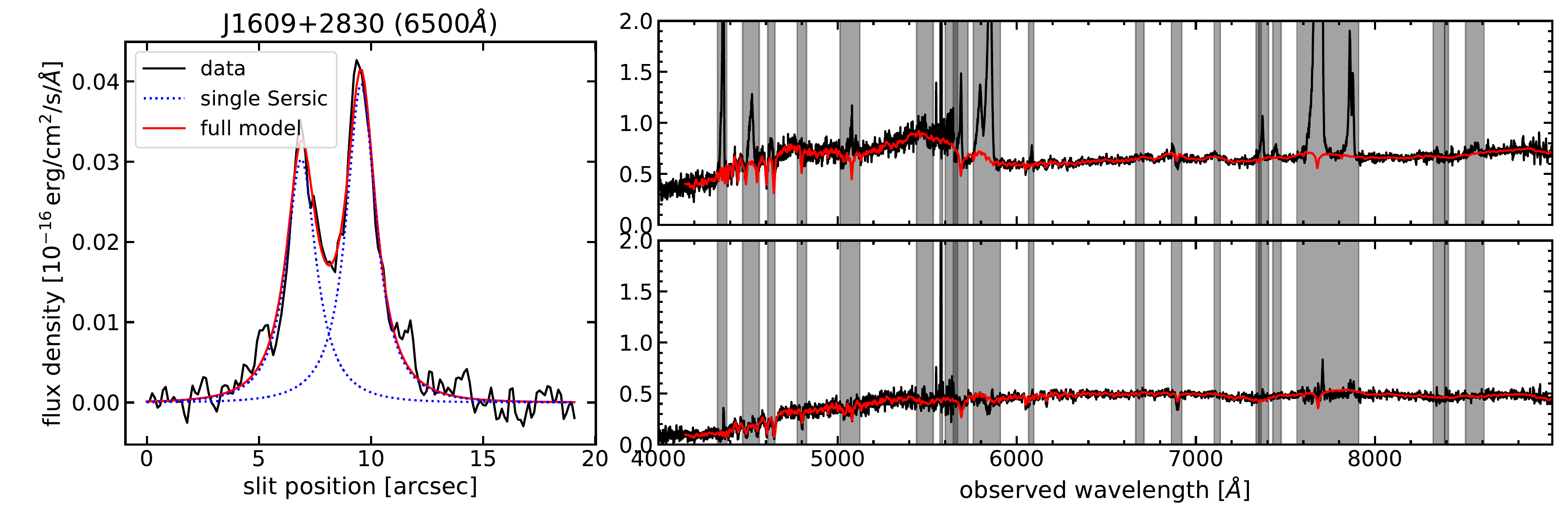}\\
 \includegraphics[width=0.9\textwidth]{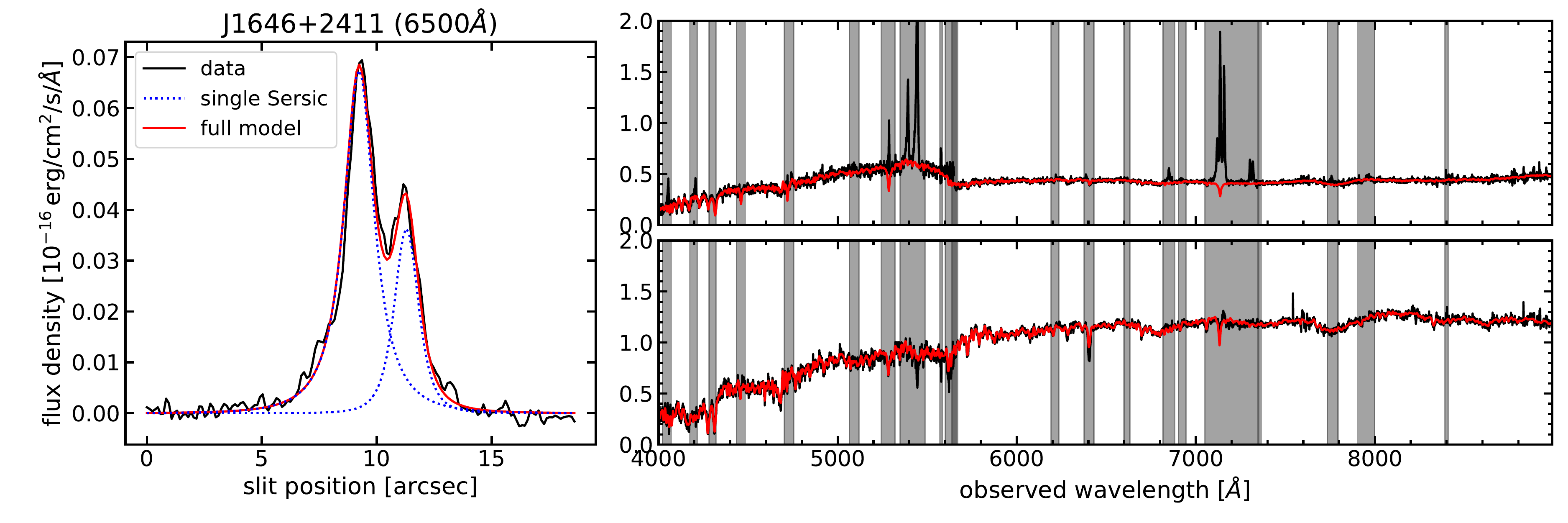}\\
 \caption{continued.}
\end{figure*}

\section{Analysis and results}
\subsection{Characterisation of the point-spread function}
As a first step in the analysis of the slit spectra, we characterise the wavelength-dependent PSF from the star observed close to the science target in the same mask observations. Again we assume a Moffat profile as described by Eq.~\ref{eq:moffat}, but we replace the radial distance $r$ with the 1D position $x-x_\mathrm{cent}$ along the slit where $x_\mathrm{cent}$ is the position of the star in the slit.

After smoothing the 2D star spectra with a 30 pixel wide median filter in the dispersion direction, we fitted each wavelength slice with a Moffat profile to obtain best-fit parameters for  $I_0$, $x_\mathrm{cent}$, $\alpha$, and $\beta$ as a function of wavelength. Afterwards, we describe the wavelength dependence of $\alpha$, $\beta$, and $x_\mathrm{cent}$ with a best-fitting fourth-order polynomial separately for the blue and red channel of the spectrograph. In Fig.~\ref{fig:PSFcharacter} we show the cross-dispersion profile of the star at 6000\,\AA\ and the measured wavelength dependence of the FWHM and $\beta$ for all the science fields. Since the blue and red channels are independent spectrographs,  their parameters are not necessarily consistent, likely due to a different spectrograph focus setup and/or optical path depending on the observations data and location on the sky. 

The PSF shape is an important characteristics of the observations because it is crucial to properly disentangle the emission of the two galaxies and galaxy nuclei. The FWHM of the PSF at 6000\AA\ is listed in Table~\ref{tab:observations} as a reference for the overall seeing during the observations. All observations except for SDSS~J0317$+$0040 are taken during good seeing conditions with $\sim$1\arcsec\ (FWHM). In the following analysis steps we always take the measured wavelength-dependence PSF for each science field into account. If the [\ion{O}{iii}] and H$\beta$ lines fall in the spectral overlap region between the two spectrograph arms, we do not combine the overlapping spectra and analyse only the data from the arm providing the highest S/N for those lines to ensure that the intrinsic PSF remains well characterised.

\begin{table*}
\caption{Cross-dispersion stellar continuum modelling and spectral parameters}\label{tab:profile_table}
\centering
\begin{tabular}{lcccccccccccc}\hline\hline
Name & $\Delta d$ & $n_1$ & $r_\mathrm{e,1}$ & $\sigma_{*,1}$ & $\mathrm{D}_{n,1}(4000)$ & $n_2$ & $r_\mathrm{e,2}$ & $\sigma_{*,2}$ & $\mathrm{D}_{n,2}(4000)$& \\
& [\arcsec] &  & [\arcsec] & [$\mathrm{km\,s}^{-1}$] &  & & [\arcsec] & [$\mathrm{km\,s}^{-1}]$ & \\\hline
SDSS J0317$+$0048 & 2.8 & $1.0\pm0.2$ & $1.1\pm0.1$ & $195\pm6$ & $1.5$ & $0.8\pm0.1$ & $1.0\pm0.1$ & $288\pm4$ & $1.8$ \\
SDSS J0805$+$2818 & 2.3 & $1.4\pm0.1$ & $0.9\pm0.1$ & $189\pm2$ & $1.3$ & $1.7\pm0.2$ & $1.8\pm0.1$ & $263\pm3$ & $1.2$ \\
SDSS J0858$+$1822 & 2.6 & $0.8\pm0.1$ & $1.3\pm0.1$ & $236\pm2$ & $1.5$ & $1.4\pm0.1$ & $1.4\pm0.0$ & $337\pm1$ & $1.7$ \\
SDSS J0947$+$6339 & 2.2 & $2.1\pm0.1$ & $2.1\pm0.1$ & $343\pm3$ & ... & $1.7\pm0.4$ & $1.0\pm0.2$ & $375\pm15$ & ... \\
SDSS J1115$+$5423 & 8.3 & $1.1\pm0.1$ & $0.5\pm0.0$ & $246\pm2$ & $2.2$ & $1.9\pm0.1$ & $1.7\pm0.1$ & $313\pm2$ & $2.0$ \\
SDSS J1609$+$2830 & 2.7 & $2.0\pm0.5$ & $1.6\pm0.2$ & $183\pm5$ & $1.9$ & $1.8\pm0.1$ & $1.7\pm0.1$ & $192\pm4$ & $1.3$ \\
SDSS J1646$+$2411 & 2.1 & $1.8\pm0.1$ & $1.7\pm0.2$ & $209\pm2$ & $1.9$ & $1.6\pm0.4$ & $0.7\pm0.2$ & $112\pm6$ & $1.5$ \\
\hline\end{tabular}
\end{table*}

\begin{table*}
\caption{Normalised peak line fluxes and kinematics of the two nuclei from LBT spectroscopy.}\label{tab:line_table}
\centering
\small
\begin{tabular}{c R{7mm} R{9mm} R{7mm} R{9mm} R{7mm} R{9mm} R{7mm} R{9mm} R{7mm} R{9mm} R{7mm} R{9mm} R{7mm} }\hline\hline
Name & $\mathrm{H}\beta_1$ & $\mathrm{[OIII]}_1$ & $\mathrm{H}\alpha_1$ & $\mathrm{[NII]}_1$ & $\sigma_1$ & $W_{\mathrm{H}\alpha,1}$ & $\mathrm{H}\beta_2$ & $\mathrm{[OIII]}_2$ & $\mathrm{H}\alpha_2$  & $\mathrm{[NII]}_2$ & $\sigma_2$ & $W_{\mathrm{H}\alpha,2}$ & $\Delta v$\\\cline{2-5}\cline{8-11} & \multicolumn{4}{c}{} & [$\mathrm{km}\,\mathrm{s}^{-1}$] & [\AA] & \multicolumn{4}{c}{} & [$\mathrm{km}\,\mathrm{s}^{-1}$] & [\AA] & [$\mathrm{km}\,\mathrm{s}^{-1}$]\\\hline
SDSS J0317$+$0048 & $0.08$ $\pm0.02$ & $1.00$ $\pm0.03$ & $1.00$ $\pm0.04$  & $0.94$ $\pm0.04$ & $194$ $\pm3$ & $47.9$ $\pm1.4$  & $0.03$ $\pm0.02$ & $0.08$ $\pm0.03$ & $0.18$ $\pm0.04$ & $0.14$ $\pm0.04$ & $214$ $\pm27$ & $8.4$ $\pm1.4$ & $321$ $\pm36$\smallskip\\
SDSS J0805$+$2818 & $0.01$ $\pm0.01$ & $0.02$ $\pm0.01$ & $0.04$ $\pm0.01$  & $0.04$ $\pm0.01$ & $180$ $\pm5$ & $6.5$ $\pm0.3$  & $0.10$ $\pm0.01$ & $1.00$ $\pm0.01$ & $1.00$ $\pm0.01$ & $0.81$ $\pm0.01$ & $320$ $\pm2$ & $169.0$ $\pm0.3$ & $34$ $\pm4$\smallskip\\
SDSS J0858$+$1822 & $0.05$ $\pm0.02$ & $0.04$ $\pm0.01$ & $0.18$ $\pm0.01$  & $0.09$ $\pm0.01$ & $190$ $\pm3$ & $15.7$ $\pm0.3$  & $0.16$ $\pm0.02$ & $1.00$ $\pm0.01$ & $1.00$ $\pm0.01$ & $1.04$ $\pm0.01$ & $207$ $\pm2$ & $88.1$ $\pm0.3$ & $30$ $\pm2$\smallskip\\
SDSS J0947$+$6339 & ... & $0.04$ $\pm0.02$ & $0.04$ $\pm0.02$ & $0.03$ $\pm0.02$ & $312$ $\pm20$ & $1.5$ $\pm0.3$ & ... & $1.00$ $\pm0.02$ & $1.00$ $\pm0.02$ & $0.62$ $\pm0.02$ & $295$ $\pm2$ & $34.8$ $\pm0.3$ & $1$ $\pm51$ \smallskip\\
SDSS J1115$+$5423 & $0.03$ $\pm0.01$ & $0.20$ $\pm0.01$ & $0.13$ $\pm0.02$  & $0.06$ $\pm0.02$ & $113$ $\pm2$ & $5.1$ $\pm0.2$  & $0.11$ $\pm0.01$ & $1.00$ $\pm0.01$ & $1.00$ $\pm0.02$ & $0.64$ $\pm0.02$ & $200$ $\pm1$ & $40.0$ $\pm0.2$ & $371$ $\pm1$\smallskip\\
SDSS J1609$+$2830 & $0.02$ $\pm0.01$ & $0.02$ $\pm0.01$ & $0.08$ $\pm0.01$  & $0.07$ $\pm0.02$ & $193$ $\pm12$ & $10.9$ $\pm0.5$  & $0.23$ $\pm0.02$ & $1.00$ $\pm0.02$ & $1.00$ $\pm0.02$ & $0.97$ $\pm0.02$ & $598$ $\pm4$ & $136.5$ $\pm1.0$ & $612$ $\pm10$\smallskip\\
SDSS J1646$+$2411 & $0.04$ $\pm0.02$ & $0.06$ $\pm0.03$ & $0.12$ $\pm0.02$  & $0.09$ $\pm0.02$ & $236$ $\pm6$ & $2.1$ $\pm0.2$  & $0.19$ $\pm0.02$ & $1.00$ $\pm0.02$ & $1.00$ $\pm0.02$ & $0.77$ $\pm0.02$ & $149$ $\pm2$ & $17.0$ $\pm0.2$ & $117$ $\pm7$\smallskip\\
\hline\end{tabular}
 \tablefoot{Emission-line flux in the blue ([\OIII], H$\beta$) and red part ([\NII], H$\alpha$) of the spectra are normalised by the [\OIII] and H$\alpha$ flux of the brighter nuclei, respectively. The implied line ratios are not corrected for Galactic extinction as we use them only for the BPT diagnostics which are insensitive to extinction.
 }
\end{table*}

\subsection{Stellar continuum modelling}
Although we are mainly interested in the emission line fluxes and ratios to study the narrow-line AGN characteristics of the two galaxy nuclei, it is crucial to subtract the stellar continuum beforehand for accurate measurements. The stellar continuum is usually modelled independently along the slit for galaxies, but our data suffer from a strong blending of the two nuclei and low S/N in the continuum already a few pixels away from the galaxy nuclei. Here we chose to obtain integrated spectra of the two galaxy components which can be properly deblended through modelling the 1D light distribution along the slit with two Sersic profiles convolved with wavelength-dependent PSF. This process ensures an accurate separation of the continuum light of both galaxies and a maximum S/N for the continuum spectra.

A single 1D Sersic profile along the slit has four free parameters, the central position in the slit ($x_\mathrm{cent}(\lambda)$), the effective radius ($r_e(\lambda)$), the Sersic index ($n(\lambda)$), and the integrated flux ($I(\lambda)$). In our case we have a superposition of two Sersic profiles with eight free parameters which are convolved with the PSF. Since the $x_\mathrm{cent}$, $r_e$, and $n$ should only vary smoothly with wavelength, we performed the fitting in three steps: 1) we fit the slit profile for each wavelength with all eight free parameters of the two Sersic model; 2) we repeat the fitting with fixed Sersic indices $n_1$ and $n_2$ at all wavelengths which are set to the median within the rest-frame wavelength range 5200--6200\AA; 3) in the final iteration, we only fit the intensities $I_1(\lambda)$ and $I_2(\lambda)$ and fix the position and radius as defined by the best-fit fourth-order polynomial to the wavelength dependence seen in the previous iteration. 

In Fig.~\ref{fig:1Dspec} we show the profile along the slit at 6500\AA\ and PSF-convolved two Sersic component model as well as the obtained spectra for each component. An optimal spectrum is extracted for each component by fitting a linear superposition of the two PSF-convolved Sersic models (with fixed parameters) to the 1D light distribution along the slit at each wavelength.  The characteristic parameters of the Sersic models are also listed in Table~\ref{tab:profile_table}. Subsequently, we model each optimally extracted spectrum with a superposition of stellar spectra from the INDO-US spectra library \citep{Valdes:2004} using PyParadise \citep[see][]{Walcher:2015,Weaver:2018,deRosa:2018}. PyParadise has the advantage that it fits the spectra after normalising, which deals much better with the systematic residuals at the wavelength of the beam splitter than adding a polynomial function to the fitting. All emission lines are masked out during fitting because their spatial distribution does not necessarily follow that of the stars, and unphysical residuals of the spatial modelling are imprinted in the spectra at their emission-line wavelengths. The best-fit continuum model is shown in Fig.~\ref{fig:1Dspec} and we also report the velocity dispersion $\sigma_*$ and the $D_n$(4000) spectral index as useful stellar age indicators \citep[e.g.][]{Bruzual:1983,Poggianti:1997,Kauffmann:2003b} in Table~\ref{tab:profile_table}.

Based on the noise-free best-fit continuum we reconstruct the 2D stellar continuum spectra using the best-fit spatial profiles as a function of wavelength. We then subtract this  2D continuum signal from the original data which leads to emission-line spectra along the slit. We remove any remaining faint continuum signal due to substructure in the galaxy light profile by running a wide median filter over 300 pixels in wavelength direction and subtract the filter continuum signal from the 2D spectra. The pure emission-line spectra are then ready for further analysis.

\begin{figure*}
 \centering
 \includegraphics[width=0.86\textwidth]{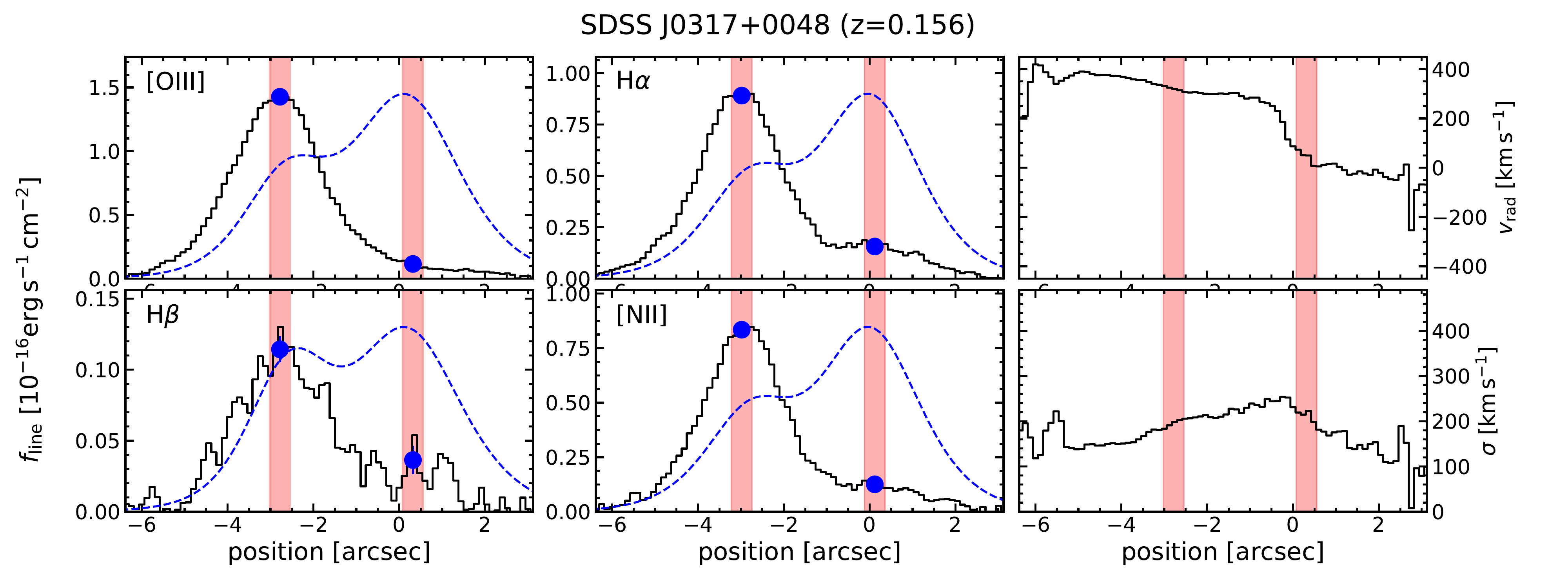}\\
 \includegraphics[width=0.86\textwidth]{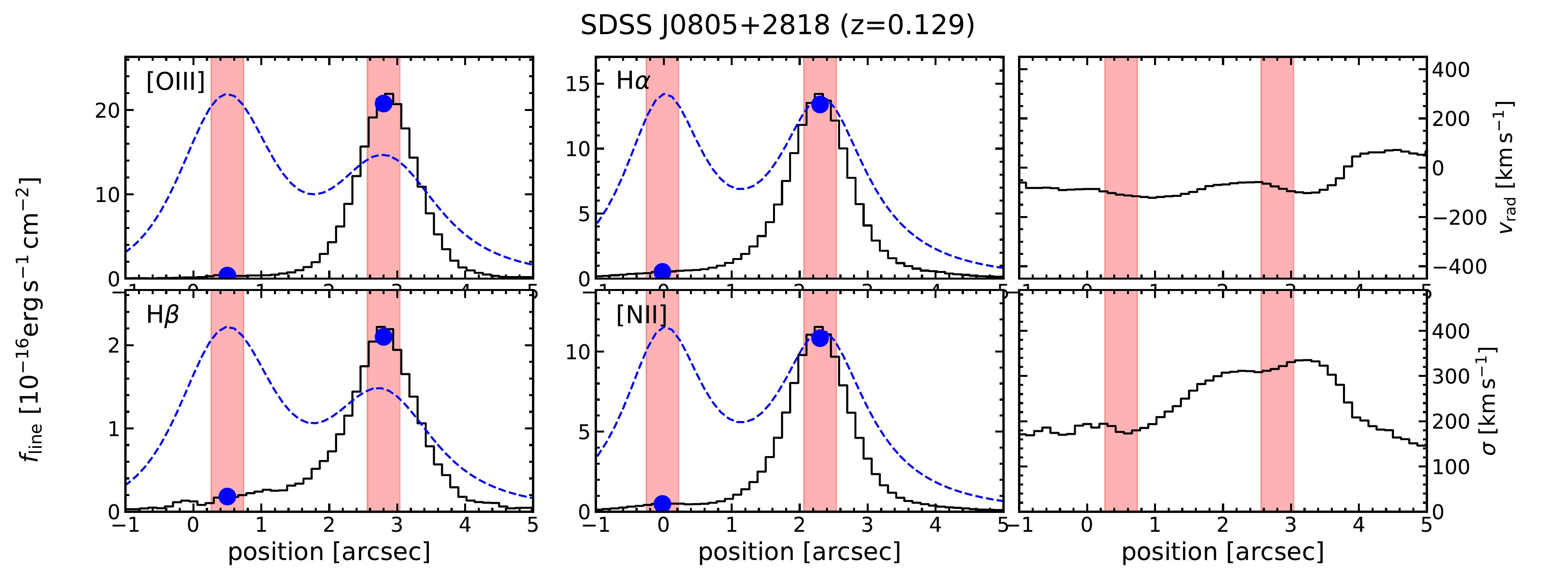}\\
 \includegraphics[width=0.86\textwidth]{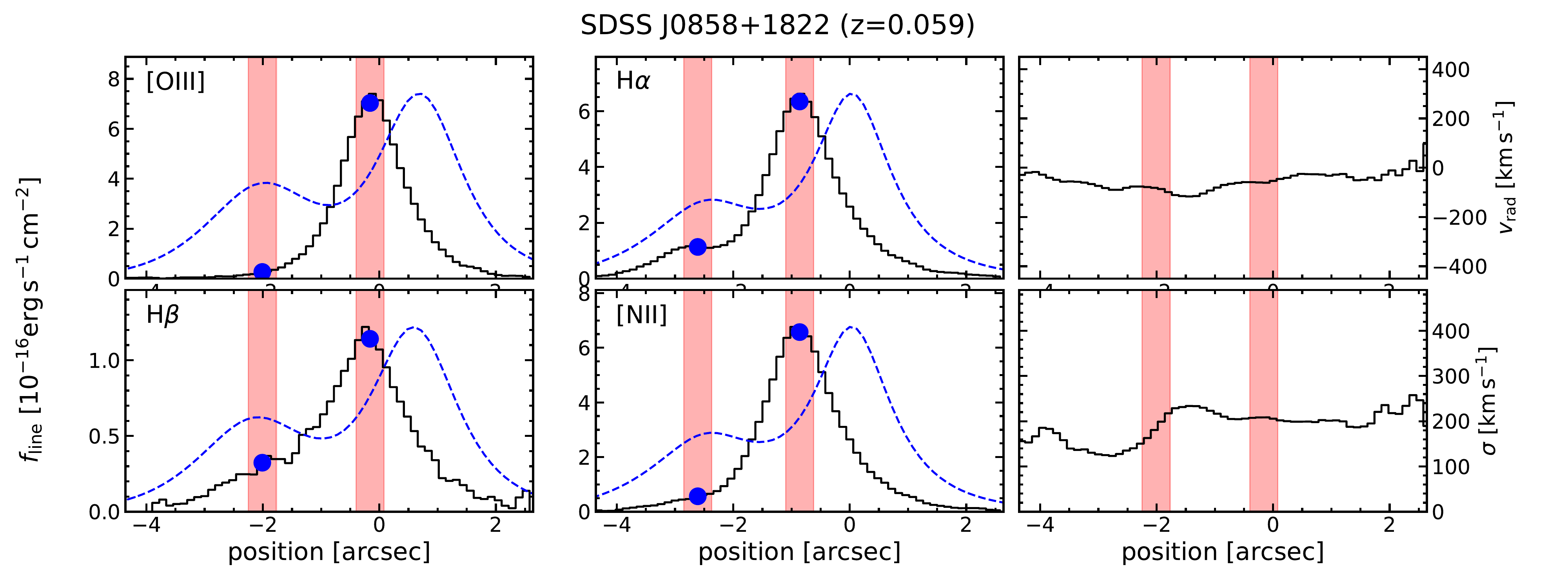}\\
 \includegraphics[width=0.86\textwidth]{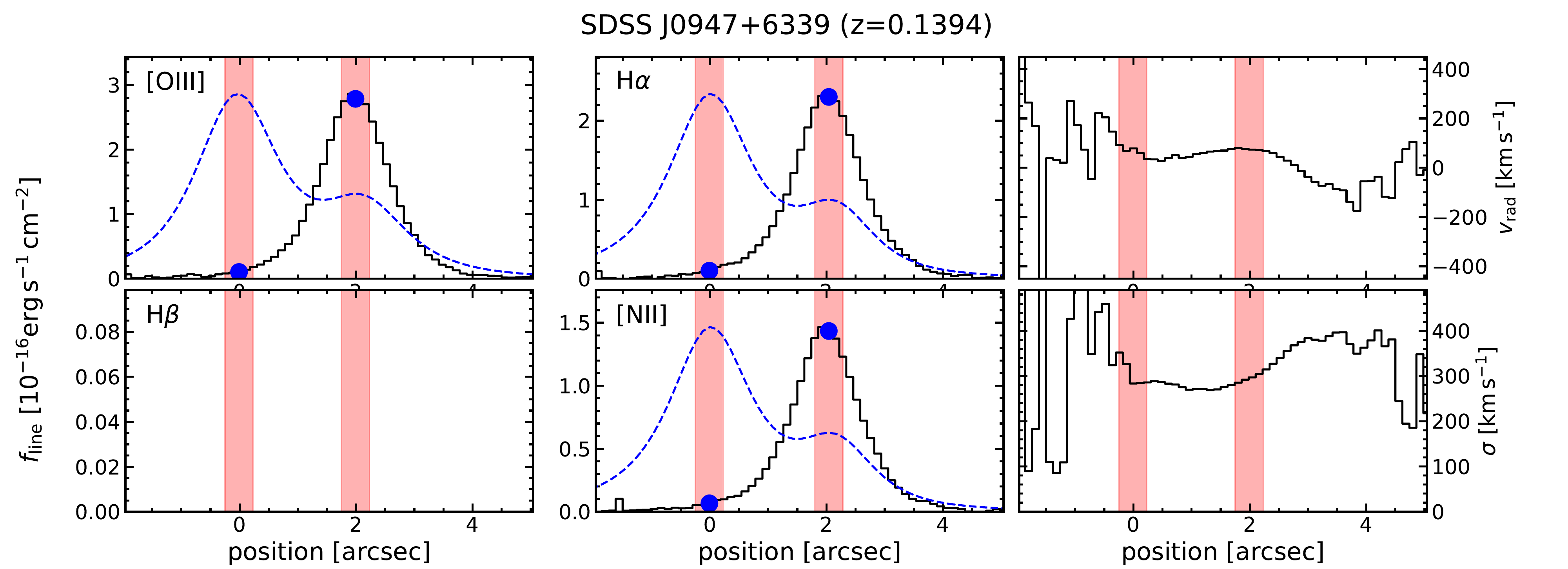}
 \caption{Overview of the emission-line fluxes and kinematics along the slit. The four left panels for each object show the 1D flux distributions for \Hb, [\OIII], \Ha, and [\NII] as black curves with the dashed blue line highlighting the continuum flux distribution for comparison. The red bands show the 0\farcs4-wide apertures used to obtain average line fluxes centred on the nuclei of the merging galaxies. The blue points represent the measured fluxes in those aperture including an error-bar based on flux variations. The two right panels present the position-velocity and the position-velocity dispersion curves for each object. Average kinematics are also measured within the same  apertures.}\label{fig:plot_lines}
\end{figure*}

\setcounter{figure}{\value{figure}-1}
\begin{figure*}
\centering
\includegraphics[width=0.86\textwidth]{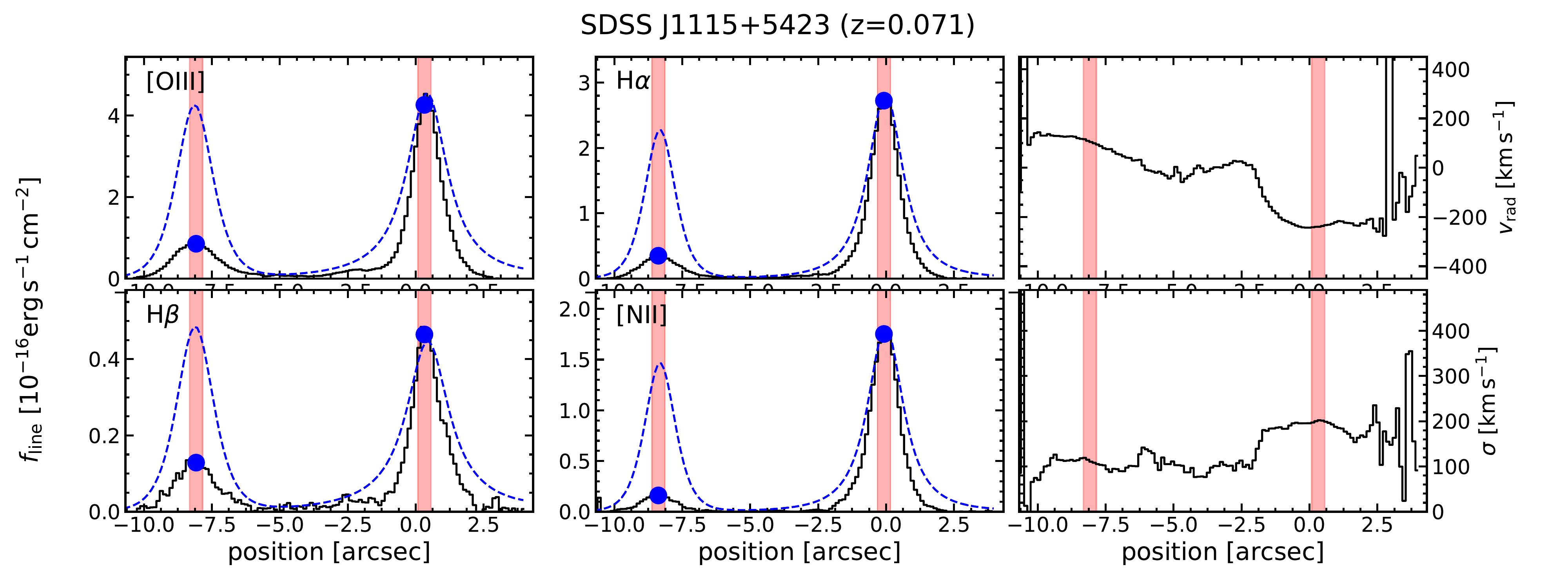}\\
\includegraphics[width=0.86\textwidth]{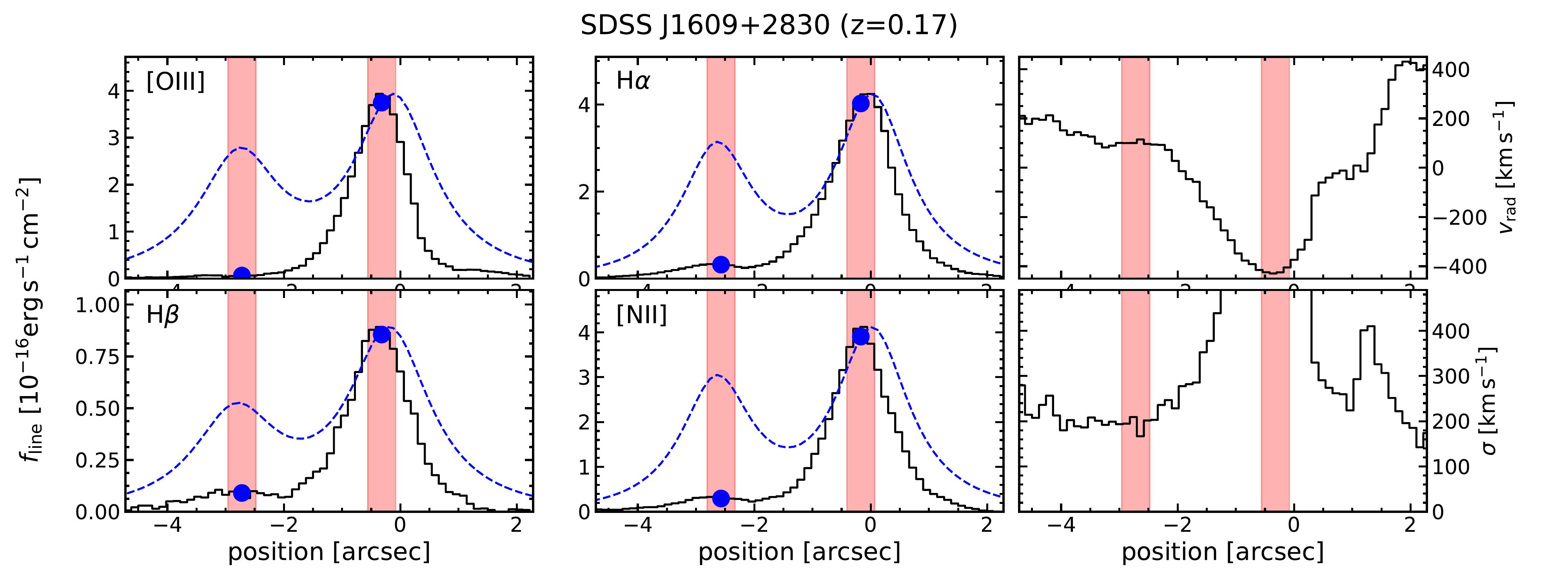}\\
\includegraphics[width=0.86\textwidth]{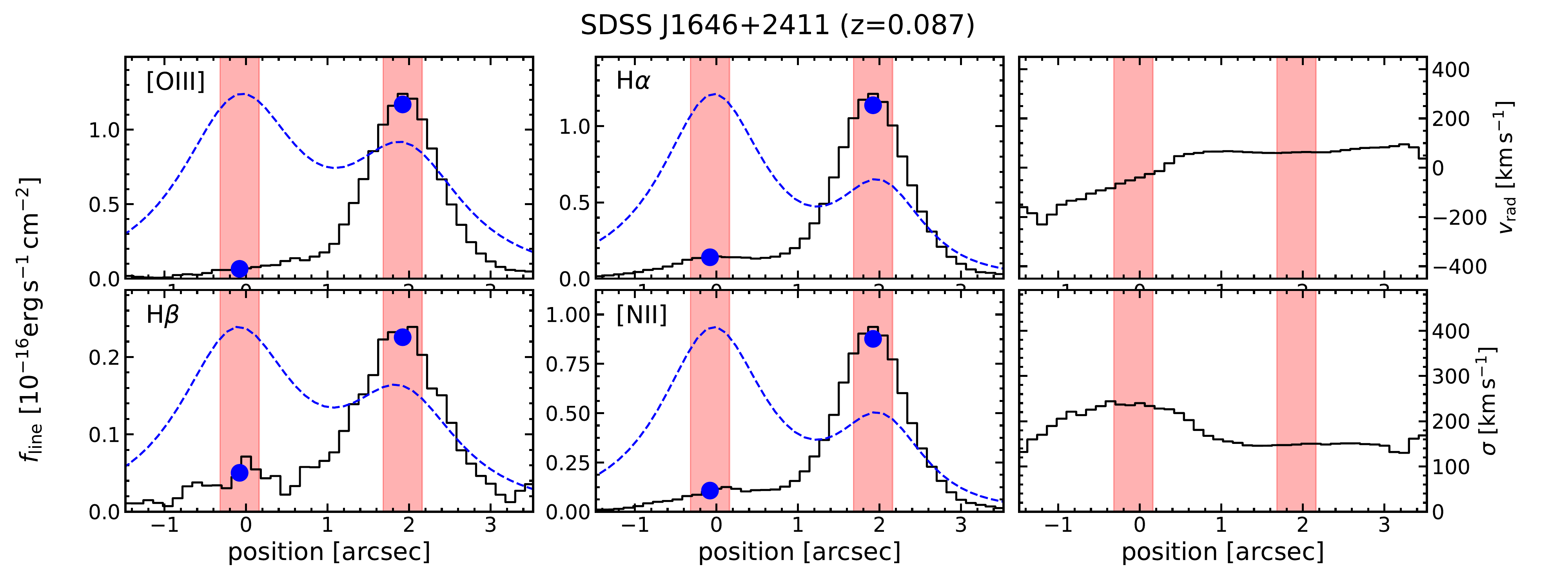}
  \caption{continued.}
\end{figure*}

\begin{figure*}
 \includegraphics[width=\textwidth]{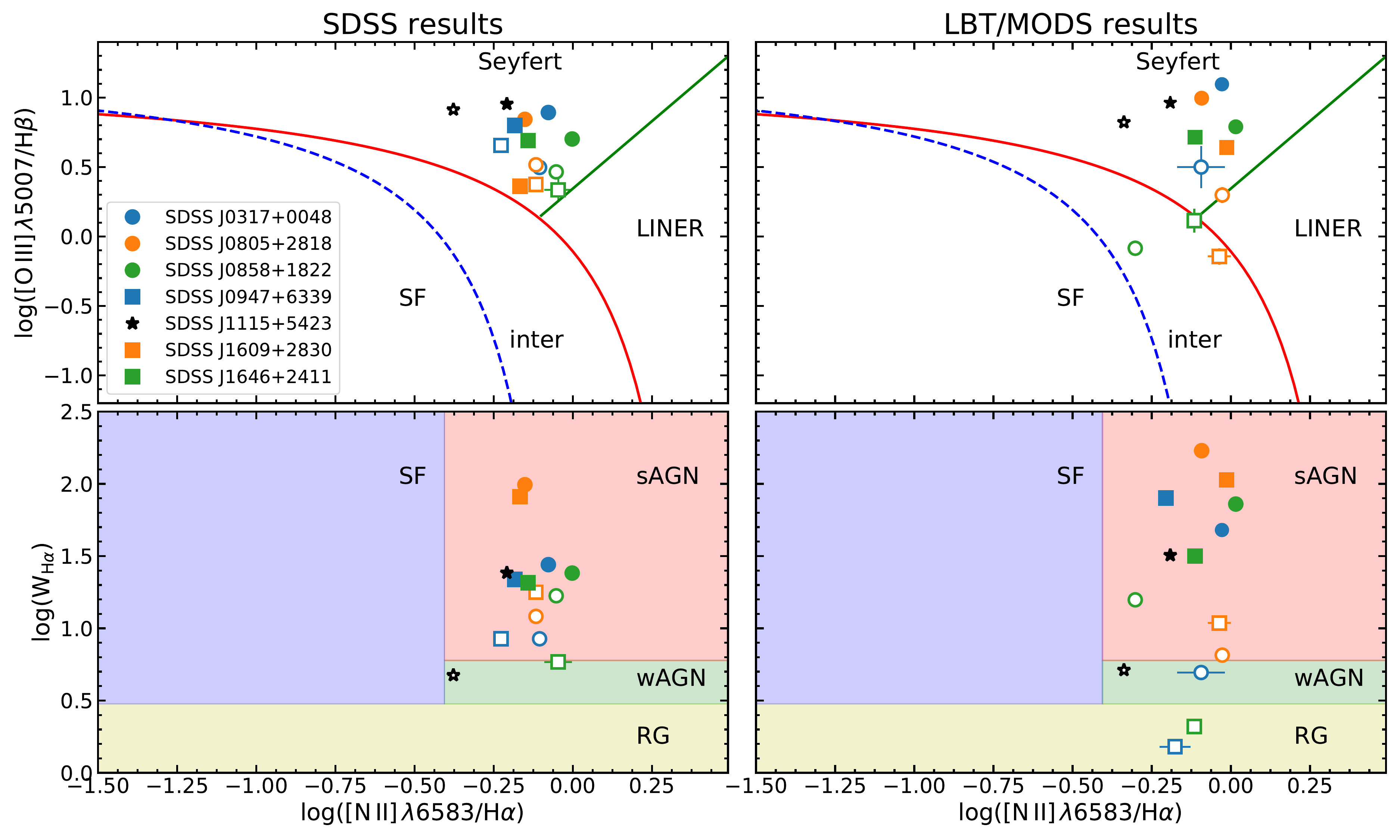}
 \caption{Comparison of emission-line diagnostics for the two nuclei from SDSS data (left panels) and LBT long slits (right panels). \textit{Top panels:} Classical BPT diagnostic diagram with demarcation lines overplotted from \citet{Kewley:2001} (red line), \citet{Kauffmann:2003} (blue line), and \citet{Stasinska:2008} (green line), which define areas of star forming galaxies (SF), Seyfert galaxies, LINERs, and intermediate (inter) galaxies. \textit{Bottom panels:} Emission-line diagnostic diagram using the equivalent width of H$\alpha$ replacing the [\ion{O}{iii}]/H$\beta$ ratio following \citet{CidFernandes:2010}. The classification scheme for SF galaxies, strong AGN (sAGN), weak AGN (wAGN), and retired galaxies (RG) is adopted from \citet{CidFernandes:2011}. The nuclei with the brighter [\ion{O}{iii}] emission line are shown with filled symbols, while the corresponding fainter nuclei are denoted with open symbols.}\label{fig:BPT}
\end{figure*}

\subsection{Emission-line measurements}
After the stellar continuum subtraction, we proceed to fit the emission lines along the slit. We used a single-Gaussian line profile for all the lines and couple their radial velocities and velocity dispersions during the fitting. The different spectral resolutions of the blue and red channel were taken into account when modelling the intrinsic velocity dispersion. The kinematical coupling drastically reduced the number of free parameters and ensures a robust flux estimates of the faintest lines. It allows us to construct line ratios in a meaningful way as the line fluxes for each emission line are emitted from the same physical association of clouds. The resulting 1D flux distributions for the \Hb, [\OIII], \Ha, and [\NII] lines along the slit are shown in Fig.~\ref{fig:plot_lines} together with the radial velocity and intrinsic velocity dispersion.

Here, we are mainly interested in the line ratios at the position of the two independent nuclei to be able to infer the true classification. Ideally, we would also fit the 1D flux distribution as a super-position of two PSF convolved functions. However, the emission line flux distribution exhibits much more substructure, unlike the smooth stellar continuum emission, and can therefore not be easily expressed with a simple analytic function. We therefore simply compute the average in line fluxes and kinematics within an aperture of 0\farcs4 around the centroid of each nucleus. The resulting measurements of peak line fluxes and the associated kinematics are listed in Table~\ref{tab:line_table}. Due to the wavelength-dependent PSF, the Balmer decrement of the peak-line fluxes cannot be robustly used to measure extinctions so we do not correct the emission lines for intrinsic dust attenuation. Therefore, we only provide relative flux with respect to [\OIII] and H$\alpha$ flux of the brightness nuclei for the blue and red part of the spectra. 

Our observations reveal that the brighter nucleus in the stellar continuum is not necessarily the more luminous emission-line source. Only in one of the seven cases (SDSS~J1609+2830) does the brighter emission-line source coincide with the brighter continuum nucleus. Surprisingly, the emission-line peak for SDSS~J0858+1822 is clearly offset by 1\arcsec\ from the position of the two apparent stellar nuclei of the ongoing merger system. This  already calls into question the initial dual AGN nature of this galaxy as the SDSS fibres were positioned at the two continuum nuclei, which means that a high fraction of the flux from the brightest emission-line source is captured by both fibres simultaneously. 

The emission-line kinematics at the position of the two nuclei are significantly different, which suggest that local emission is dominating even for the secondary nuclei. An exception is SDSS~J0947$+$6339, where the radial velocity and velocity dispersion are consistent within the errors. This means that the PSF-smeared emission of the primary may even dominate the emission at the location of the second nucleus. This should be taken into account when interpreting the emission-line ratios. The emission lines are very broad for the primary AGN nucleus in SDSS~J1609$+$2830 with a velocity dispersion of 600\,km/s and a blue shift of more than 600\,km/s with respect to the secondary nucleus. This suggests that a powerful outflow is released by the AGN which is interesting by itself, but outside the scope of this paper.

\subsection{Emission-line diagnostics for the nuclei}

Based on the obtained emission line measurements from the original SDSS spectra and the LBT long-slit analysis presented here, we construct two emission-line diagnostic diagrams (see Fig.~\ref{fig:BPT}). The first  is the classical [\OIII]/\Hb\ versus [\NII]/\Ha\ emission-line diagnostics diagram \citep[BPT diagram,][]{Baldwin:1981,Veilleux:1987} and the second  replaces the [\ion{O}{iii}]/H$\beta$ ratio by the equivalent width of H$\alpha$ ($W_{\mathrm{H}\alpha}$), as proposed by \citet{CidFernandes:2010}, the so-called WHAN diagram. The Seyfert-Seyfert pre-selection from SDSS of the two nuclei  is confirmed by the BPT diagram shown in Fig.~\ref{fig:BPT} (upper left panel), although the second nucleus of SDSS~J1115$+$5423 and SDSS~J1646+2411 is in the weak AGN regime following the WHAN diagram. Hence, all seven systems have  been assumed as dual AGN in interacting galaxy systems  based on these optical diagnostics.

However, the MODS spatially resolved spectroscopy reveals a significantly different picture, as shown in the right panels of Fig.~\ref{fig:BPT}. Except for SDSS~J1115+5423,  all the fainter secondary nuclei, which were initially classified clearly as Seyfert-like AGN, are changing their apparent emission-line classification. The secondary nuclei of SDSS~J0947$+$6339 and SDSS~J1646$+$2411 clearly fall into a lower ionisation regime which are likely ionised by the stellar population instead of an AGN given the low equivalent width of H$\alpha$ in the WHAN classification. SDSS~J0858$+$1822 is the system where the peak in the [\ion{O}{iii}] emission is located between the two continuum nuclei making the dual AGN signature of the SDSS fibre obsolete, and shows that one of the nuclei is dominated by star formation, as confirmed by the BPT diagram, which cannot be properly distinguished in the WHAN diagram in the transition zone. 
While SDSS~J0317$+$0048 and SDSS~J0805$+$2818 remain in the Seyfert/AGN classification in the BPT, the equivalent width of H$\alpha$ is very low and certainly in the weak AGN regime. Only SDSS~J1609$+$2830 seems to retain an unambiguous dual AGN classification although the BPT diagnostic reveals a significant SF or LINER-like contribution to the emission inconsistent with the previous Seyfert-Seyfert classification based solely on SDSS.

\begin{figure*}
 \includegraphics[width=\textwidth]{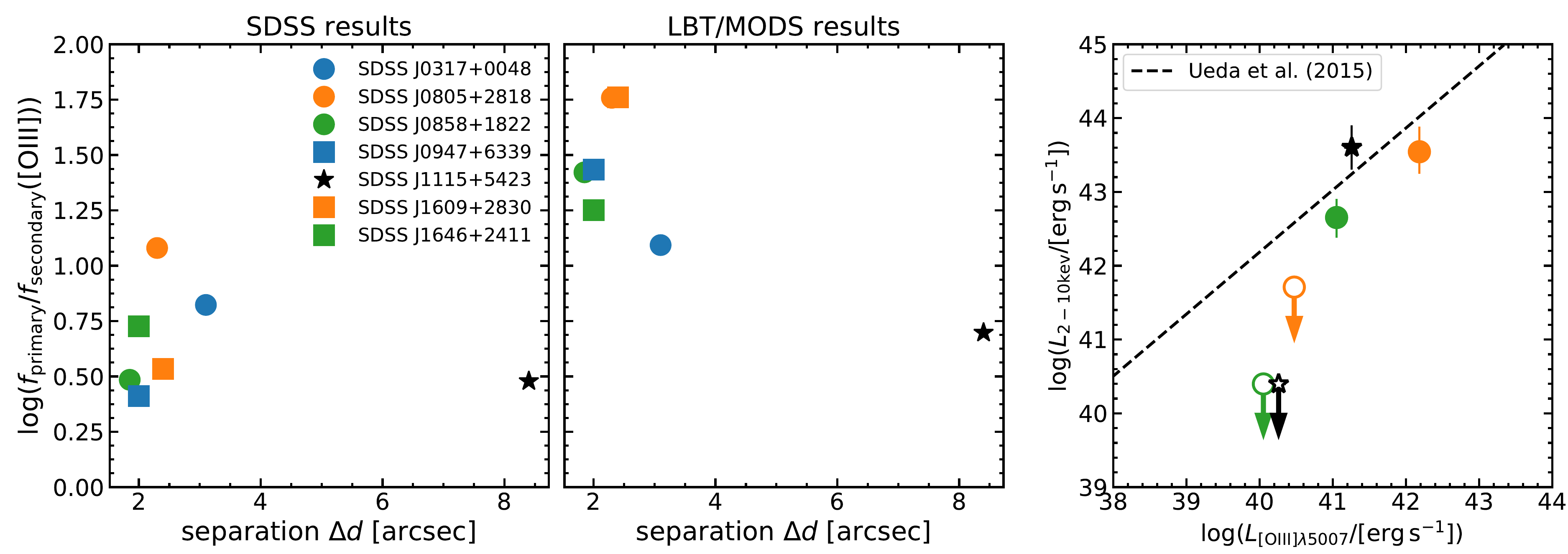}
 \caption{[\ion{O}{iii}] flux ratios based on SDSS (left panel) and LBT (middle panel) spectroscopy for the two nuclei as a function of separation. The 2--10\,keV X-ray luminosities based on \textit{Chandra} observations are compared to the [\ion{O}{iii}] luminosities for three of the sources in the right panel. The secondary nuclei are undetected in X-rays and $5\sigma$ upper limits on the X-ray luminosity are provided in these cases and shown as unfilled symbols. The dashed line shows the correlation determined by \citet{Ueda:2015} for a large sample of obscured (type 2) AGN as a reference.}\label{fig:O3_AGN_lum}
\end{figure*}

\section{Discussion}
\subsection{Interpretation of dual AGN signatures in very close systems from SDSS}
A key result of this work is that nearly all of the six putative AGN systems with separations of less than 3\arcsec\ change their initial Seyfert--Seyfert classification from the SDSS fibre spectra. One of the targets  turns out to be completely offset with respect to the fibre positions targeting  optically bright continuum knots. This  means that the unambiguous (Seyfert 2--Seyfert 2) dual AGN fraction at the closest separations ($<$3\arcsec) reported by \citet{Liu:2011} is overestimated by a larger factor. However, given the complex selection function and completeness of the SDSS for obtaining spectra so close together, which avoids merging or neighbouring galaxies, we do not attempt to  extrapolate our findings on the dual AGN fraction from our sample to the overall population.

Another big complication is the actual classification of galaxies based solely on the emission-line diagnostic diagrams which are often ambiguous to clearly identify AGN signatures. In particular, the LINER-like emission-line ratios may not be necessarily produced by AGN and are shown to be often produced by post-AGB stars in galaxies with old stellar populations \citep[e.g.][]{CidFernandes:2011,Singh:2013}. Also emission-line ratios in the intermediate region between the star forming and AGN classification could be entirely powered by a starburst or an actual mix of star formation and AGN photo-ionisation. In both cases the AGN signature of the corresponding nucleus needs to be verified at other wavelengths, for example radio \citep[e.g.][]{Bondi:2010,Fu:2015,MuellerSanchez:2015,Bondi:2016} or X-rays \citep[e.g.][]{Koss:2012,Comerford:2015,Ellison:2017,Hou:2019}. For LINERs it may be more likely to detect the radio emission of a jet if the accretion disc is  in a radiatively inefficient mode. 

While it has  often been highlighted that double-peaked [\OIII] emitters require spatially resolved spectroscopy and multi-wavelength follow-up to confirm the dual AGN nature, our study reveals that  spatially resolved spectroscopy is also indispensable for verifying the dual AGN at separations of $<5\arcsec$ even if the separate spectra were taken with fibre spectroscopy such as SDSS, LAMOST, and 4MOST \citep{deJong:2019}, or WEAVE \citep{Dalton:2014} in the near future. While we only studied the spillover effect for obscured dual AGN candidates, we note that the same effect will also impact unobscured dual AGN candidates at close separation. In that case, the light of broad emission lines would also  be detected in fibres placed a few arcsec away depending on brightness and seeing conditions. While the detection of broad Balmer lines is a much less questionable AGN signature than the emission-line ratios, it may suffer from the same spillover effects and also requires spatially resolved spectroscopy for confirmation.

\subsection{X-ray--to--[\OIII] line ratios in close dual AGN candidates}
A systematic study of X-ray follow-up observations with Chandra was presented by \citet{Hou:2019} for five of the compact dual AGN candidates selected from SDSS, including SDSS~J0805+2818. Only two of the five putative dual AGN candidates could be confirmed with clear X-ray detection in both nuclei despite the unobscured AGN classifications from the SDSS spectroscopy. \citet{Hou:2019} interpret their results as an indication for systematically  lower X-ray--to--[\ion{O}{iii}] luminosity ratios in those systems. Based on our analysis, we suggest that the measured [\ion{O}{iii}] luminosities are strongly biased in the SDSS spectra. As we  show in Fig.~\ref{fig:O3_AGN_lum}, the [\ion{O}{iii}] luminosity can be overestimated by an order of magnitude for the secondary nucleus at $<$3\arcsec\ separations. 

\textit{Chandra} X-ray observations are available for SDSS~J0805$+$2818 \citep{Barrows:2016,Hou:2019}, SDSS~J0858$+$1822 (Proposal ID: 14700279, PI: Liu, unpublished), and SDSS~J1115$+$5423 \citep{Barrows:2016}. In all Chandra images the primary nuclei are always detected, while the secondary nuclei are usually undetected. We derived a $5\sigma$ upper limit on the 2--10\,keV flux based on the background noise, which corresponds to $<5\times10^{41}\,\mathrm{erg\,s}^{-1}$, $<2.2\times10^{40}\,\mathrm{erg\,s}^{-1}$, and $<2.5\times10^{40}\,\mathrm{erg\,s}^{-1}$ for the three sources, respectively, adopting a $N_H=10^{22}\mathrm{cm}^{-2}$ and a power-law index of $\Gamma=1.7$. The X-ray fluxes of the primary nuclei for SDSS~J0805$+$2818 and SDSS~J1115$+$5423 are published in \citet{Barrows:2016}, but we re-visit the spectra and adjust luminosity estimates based on spectral shape.  For SDSS~J0858$+$1822 we found that the bulk of X-ray emission is located  between the optical nuclei in the SDSS images, which is fully consistent with our LBT spectroscopy. Furthermore, the X-ray spectra of SDSS~J0805$+$2818 and SDSS~J0858$+$1822 are rather soft, which could be due to a heavy obscuration of the primary nuclear component up to 4--5\,keV (and potentially up to the Compton-thick level of 10\,keV when the spectrum is dominated by the disc reflected component) and leaving a soft X-ray scattered component dominating at low energies.

At the S/N of the X-ray spectra a detailed modelling of the reflection spectrum is difficult. To obtain the upper limit to the X-ray luminosity, we assume that the direct X-ray radiation is totally absorbed by a Compton-thick medium ($N_H >10^{24}\mathrm{cm}^{-2}$) and that we are observing the reflected component in the 2--10\,keV energy band.  We apply a mean correction factor of 70 to obtain the intrinsic 2--10\,keV luminosity \citep[e.g.][]{Lamastra:2009,Marinucci:2012}, which is close to an absolute upper limit for the intrinsic X-ray luminosity.

In Fig.~\ref{fig:O3_AGN_lum} (right panel) we compare the 2--10\,keV X-ray luminosities against the [\ion{O}{iii}] luminosities. For the secondary nuclei we adopt the [\ion{O}{iii}] flux ratios from the LBT data and applied it to the flux measured for the primary by SDSS. This avoids potential aperture losses in the slit spectra for the absolute luminosities. We see that the primary nuclei closely follow  the X-ray--to--[\ion{O}{iii}] correlation by \citet{Ueda:2015}, but only if the correction factors are applied for the reflection-dominated spectra. The secondary nuclei fall short of correlation by 2 orders of magnitude even if a correction factor is applied to the upper limits for reflection except for SDSS~J0805$+$1822. Hence our data is consistent with normal X-ray--to--[\ion{O}{iii}] ratios considering the updated emission-line ratio diagnostics which show that the secondary nuclei are not necessarily genuine obscured type 2 AGN. The interpretation  of exceptional X-ray--to--[\ion{O}{iii}] ratios by \citet{Hou:2019} was based on the SDSS line fluxes which simply need to be significantly revised. SDSS~J0805$+$1822 could be a dual AGN consistently inferred from the emission lines, but the corresponding X-ray source is quite faint and  only consistent with an AGN interpretation if the X-ray luminosity is scaled up by a factor of 70 for a Compton-thick source. The X-ray observations of SDSS~J0858$+$1822 also confirm the non-AGN nature of the secondary nucleus which appears to be dominated by star formation in the LBT spectra. SDSS~1115$+$5432 shows AGN ionised gas along the entire slit between the galaxies, which is a clear indication that the ionisation cone of the primary is pointing towards the companion galaxy. This opens the possibility that the ionised gas in the secondary neighbouring galaxy is actually illuminated by the primary AGN \citep{Xu:2009} as the narrow-line region is known to extend out to scales of  several kiloparsec   at this AGN luminosity \citep[e.g.][]{Hainline:2013,Husemann:2014}. This may also explain the low equivalent width of H$\alpha$ for the secondary nucleus in SDSS~J1115$+$5423 despite the clear Seyfert-like emission-line ratios. Only 3D spectroscopy would allow us to map the entire ENLR structure.

\subsection{AGN triggering in interacting systems}
\begin{figure}
 \resizebox{\hsize}{!}{\includegraphics{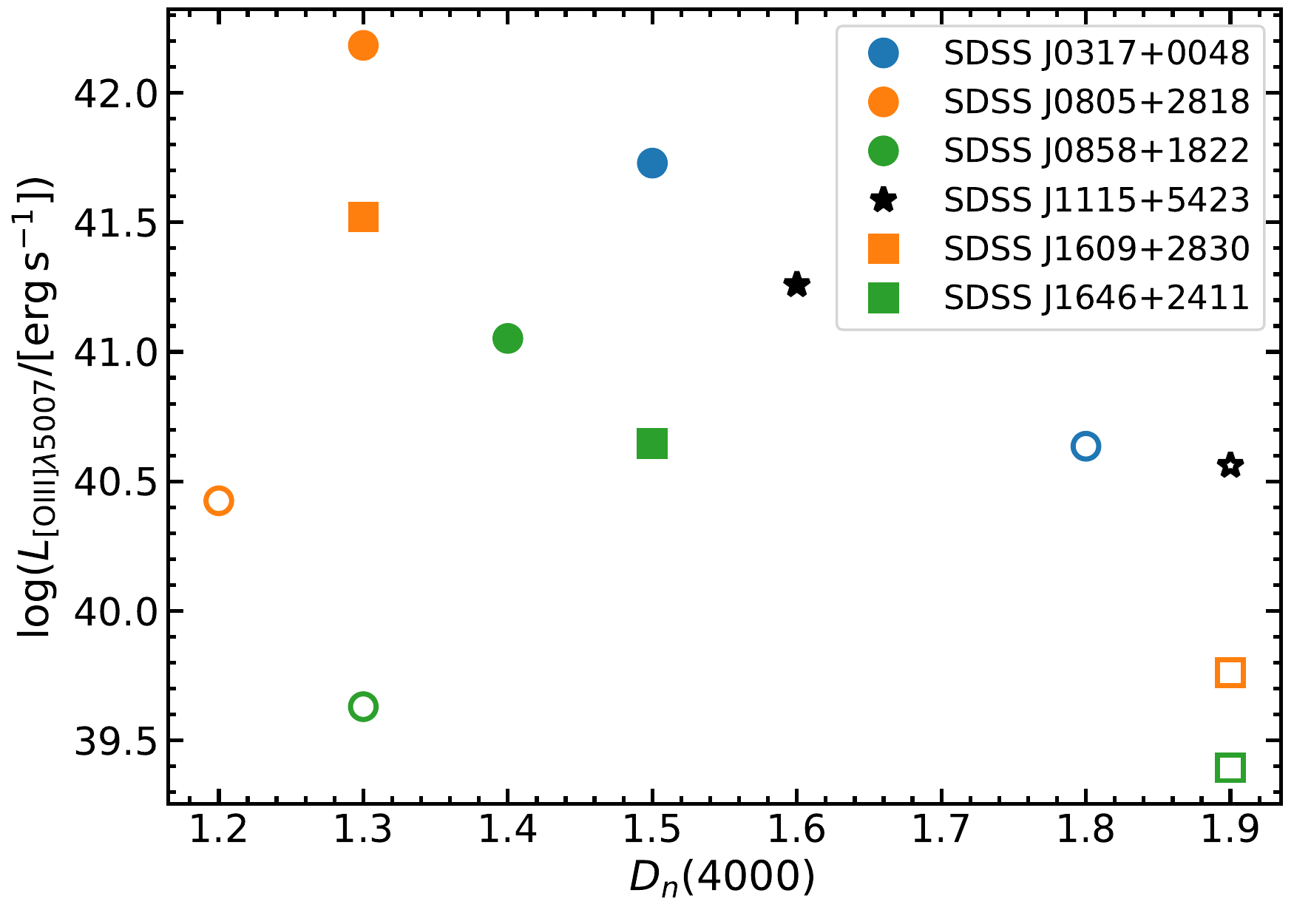}}
 \caption{[\ion{O}{iii}] luminosity against $D_n(4000)$ stellar index as a stellar population age indicator. Filled symbols correspond to the primary AGN where the [\ion{O}{iii}] luminosity is taken from the SDSS spectrum and the open symbol correspond to the secondary nuclei for which the luminosity is called according to the flux ratio measured by our LBT spectra.}\label{fig:stellarpop}
\end{figure}

While it is difficult to explain the triggering of AGN, it has been suggested that radiatively efficient AGN are often associated with recent star formation activity in their host galaxies \citep[e.g.][]{Canalizo:2000,Gonzalez-Delgado:2001,Kauffmann:2003,Davies:2007,Wild:2010}. One potential explanation for this  connection is the common gas reservoir necessary for the two processes \citep[e.g.][]{Maiolino:2007,Netzer:2009,Rosario:2012,Mullaney:2012,Chen:2013,Hickox:2014,Husemann:2018}. We therefore explore whether the nuclear activity in our interacting galaxies is linked to the star formation history in their host galaxies. Here we use the $D_n(4000)$ stellar index as a luminosity-weighted age indicator of the stellar population \citep[e.g.][]{Hamilton:1985,Poggianti:1997,Kauffmann:2003b}. Values $D_n(4000)>1.5$ correspond to stellar ages of $t_\mathrm{age}>1$\,Gyr and $D_n(4000)<1.5$ to $t_\mathrm{age}<1$\,Gyr adopting solar metallicities. In Fig.~\ref{fig:stellarpop}, we compare the [\ion{O}{iii}] luminosity with the $D_n(4000)$ index. 

We find that all primary nuclei with unambiguous signatures of an obscured Seyfert-like AGN are associated with very young stellar populations indicating recent or ongoing star formation activity. The secondary nuclei with ambiguous ionisation classifications exhibit significantly older stellar populations sometimes close to passive galaxies. The only exceptions are the secondary nuclei of SDSS~J0805$+$2818 and SDSS~J0858+1822, which can be explained by a weak, heavily obscured AGN in the first case and a star formation dominated region in the second case. On the other hand, the old stellar populations for the other secondary nuclei support the notion that post-AGB stars or shock-ionisation are likely the primary reasons for the gas excitation and not necessarily an embedded AGN. Overall, the stellar ages support the diversity of the emission-line diagnostics from the LBT spectra which challenge the original classification of dual AGN system from SDSS spectroscopy.

\section{Conclusions}
In this paper we analysed follow-up spatially resolved slit spectroscopy taken with MODS at the LBT of seven dual candidates with separations of a few arcsec previously identified in SDSS by \citet{Liu:2011}. Classical emission-line diagnostics based on the SDSS spectra suggest unambiguous obscured Seyfert-Seyfert pair classification in all cases, but our slit spectra reveal a much more diverse nature of the activity in the nuclei of the ongoing merging systems. We attribute the diagnostic difference between the SDSS and LBT observations due to the relatively large SDSS fibres. The spectra of the secondary nuclei are contaminated by spillover light from the primary nuclei due to the seeing at the time of observations and small separation of the nuclei. Considering multiple emission-line diagnostic diagrams together with age indicators of the stellar population shows that nearly all of the secondary nuclei are likely not AGN, but rather consistent with pure star formation or LINER-like emission potentially powered by post-AGB stars \citep[e.g.][]{Singh:2013}. 

X-ray follow-up observations usually identified only one AGN in such SDSS-selected interacting systems, even for the large-separation source SDSS~J1115$+$5423. Based on the original dual AGN classification the low X-ray fluxes for the secondary nuclei have been interpreted as a systematically lower X-ray--to--[\ion{O}{iii}] ratio in interacting systems.  We propose an alternative explanation that the secondary nuclei are not necessarily AGN, and therefore actually do not need to follow the classical X-ray--to--[\ion{O}{iii}] correlations of AGN. In addition, heavy obscuration of the primary nuclei could also play a significant role in the interpretation of the ratio which require deep X-ray spectra to correct properly. 

This study highlights the need for spatially resolved spectroscopy of such compact interacting galaxy systems to properly classify the activity in the independent nuclei. Hence, special care needs to be taken when interpreting fibre-based spectroscopy of small separation sources due to seeing effects, which also remains an issue  for the ongoing and upcoming fibre-based surveys such as LAMOST, WEAVE, and 4MOST. In addition, a multi-wavelength approach is necessary to verify the activity types that are  often ambiguous, even from high-quality rest-frame optical spectroscopy.

\begin{acknowledgements}
We thank the anonymous referee for constructive feedback that improved the quality and clarity of the presented work. BH would like to thanks the European Space Agency (ESA) for the hospitality and financial support for a science visit at ESA/ESTEC in Noordwijk (Netherlands)  part of this work has been conducted.

SB and ADR acknowledges financial support from the Italian Space Agency (ASI) under grant ASI-INAF 2017-14-H.O. CV is grateful for financial support from ASI under the contracts ASI-INAF I/037/12/0 and ASI-INAF n.2017-14-H.0.

This paper used data obtained with the MODS spectrographs built with funding from NSF grant AST-9987045 and the NSF Telescope System Instrumentation Program (TSIP), with additional funds from the Ohio Board of Regents and the Ohio State University Office of Research.

All  authors are  members  of  the MAGNA team   (www.issibern.ch/teams/agnactivity/Home.html) and acknowledge support from the International Space Science Institute (ISSI) in Bern, Switzerland.

Funding for SDSS-III has been provided by the Alfred P. Sloan Foundation, the Participating Institutions, the National Science Foundation, and the U.S. Department of Energy Office of Science. The SDSS-III web site is http://www.sdss3.org/. SDSS-III is managed by the Astrophysical Research Consortium for the Participating Institutions of the SDSS-III Collaboration including the University of Arizona, the Brazilian Participation Group, Brookhaven National Laboratory, Carnegie Mellon University, University of Florida, the French Participation Group, the German Participation Group, Harvard University, the Instituto de Astrofisica de Canarias, the Michigan State/Notre Dame/JINA Participation Group, Johns Hopkins University, Lawrence Berkeley National Laboratory, Max Planck Institute for Astrophysics, Max Planck Institute for Extraterrestrial Physics, New Mexico State University, New York University, Ohio State University, Pennsylvania State University, University of Portsmouth, Princeton University, the Spanish Participation Group, University of Tokyo, University of Utah, Vanderbilt University, University of Virginia, University of Washington, and Yale University. 
\end{acknowledgements}

\bibliographystyle{aa}
\bibliography{references}
\end{document}